%% file: main.tex
\documentclass[sigconf,screen]{acmart}
\setcopyright{none} 
\acmConference[CCS'23]{30th ACM Conference on Computer and Communications Security}{Nov 26--30, 2023}{Copenhagen, Denmark}
\acmYear{2023}

\usepackage{graphicx}
\usepackage{listings}
\usepackage{float}
\usepackage{multirow}
\usepackage{booktabs}
\usepackage[scaled]{helvet}
\usepackage[linesnumbered,ruled,lined,noend]{algorithm2e}
\usepackage{mathrsfs}
\usepackage{mathpartir}
\usepackage{dsfont} 
\usepackage{stmaryrd}
\usepackage{url}
\usepackage{textcomp} 
\usepackage{mathpartir} 
\usepackage{caption}
\usepackage{multicol}
\usepackage{blindtext}
\usepackage{alltt} 
\usepackage{bbm}
\usepackage{alltt}
\usepackage{verbdef}
\usepackage{xspace}
\usepackage{verbatim}
\usepackage[inline, shortlabels]{enumitem}
\usepackage{lipsum}
\usepackage{wrapfig}
\usepackage{xcolor}
\usepackage{hyperref}
\hypersetup{linkcolor=black,citecolor=black,urlcolor=RubineRed}

\usepackage{enumitem}
\usepackage{fancyvrb}
\usepackage{pifont}
\usepackage{enumitem}
\usepackage[frozencache,cachedir=.]{minted}
\usepackage{tcolorbox}
\usepackage{shortcuts}
\usepackage{flushend}
\usepackage{threeparttable}

\input{defs}

\setlist[itemize]{leftmargin=*}
\setlist[enumerate]{leftmargin=*}

\DeclareRobustCommand{\ttfamily}{\fontfamily{lmtt}\selectfont}

\usepackage{xcolor}
\hypersetup{colorlinks,
  linkcolor=ACMDarkBlue,
  citecolor=ACMPurple,
  urlcolor=ACMDarkBlue,
  filecolor=ACMDarkBlue}

\floatname{algorithm}{Algorithm}

\begin{document}

\title{Greybox Fuzzing of Distributed Systems}

\author{Ruijie Meng}
\authornote{Joint first authors}
\affiliation{%
  \institution{National University of Singapore}
  \country{Singapore}
  }
\email{ruijie_meng@u.nus.edu}

\author{George P{\^i}rlea}
\authornotemark[1]
\affiliation{%
  \institution{National University of Singapore}
  \country{Singapore}
  }
\email{gpirlea@comp.nus.edu.sg}

\author{Abhik Roychoudhury}
\authornote{Corresponding author}
\affiliation{%
  \institution{National University of Singapore}
  \country{Singapore}
  }
 \email{abhik@comp.nus.edu.sg}

\author{Ilya Sergey}
\affiliation{%
  \institution{National University of Singapore}
  \country{Singapore}
  }
\email{ilya@nus.edu.sg}

\begin{abstract}
\input{abstract}

\end{abstract}

\maketitle

\input{intro}

\input{overview}

\input{design}
\input{evaluation}

\input{related}

\input{conclusion}
\begin{acks}

We would like to sincerely thank all the anonymous reviewers and our shepherd for their valuable feedback that greatly helped us to improve this paper. 
This research is supported by the National Research Foundation, Singapore, and Cyber Security Agency of Singapore under its National Cybersecurity R\&D Programme (Fuzz Testing <NRF-NCR25-Fuzz-0001>), and also by 
Sergey's Amazon Research Award. Any opinions, findings and conclusions, or recommendations expressed in this material are those of the author(s) and do not reflect the views of National Research Foundation, Singapore, and Cyber Security Agency of Singapore.

\end{acks}


\bibliographystyle{ACM-Reference-Format}
\bibliography{references}


\end{document}

%% file: defs.tex
\theoremstyle{remark}

\newcommand{\isep}{\mathrel{{.}\,{.}}\nobreak}
\newcommand{\var}{\text}
\newcommand{\fun}{\text}

\makeatletter 
\def\arcr{\@arraycr}
\makeatother

\definecolor{deepblue}{rgb}{0,0,0.5}
\definecolor{deepred}{rgb}{0.6,0,0}
\definecolor{deepgreen}{rgb}{0,0.5,0}
\definecolor{halfgray}{gray}{0.55}
\definecolor{ipythonframe}{RGB}{207, 207, 207}

\definecolor{a}{HTML}{D7191C}
\definecolor{b}{HTML}{FDAE61}
\definecolor{c}{HTML}{ABDDA4}
\definecolor{d}{HTML}{2B83BA}
\definecolor{d}{HTML}{2B23B2}

\definecolor[named]{ACMBlue}{cmyk}{1,0.1,0,0.1}
\definecolor[named]{ACMYellow}{cmyk}{0,0.16,1,0}
\definecolor[named]{ACMOrange}{cmyk}{0,0.42,1,0.01}
\definecolor[named]{ACMRed}{cmyk}{0,0.90,0.86,0}
\definecolor[named]{ACMLightBlue}{cmyk}{0.49,0.01,0,0}
\definecolor[named]{ACMGreen}{cmyk}{0.20,0,1,0.19}
\definecolor[named]{ACMPurple}{cmyk}{0.55,1,0,0.15}
\definecolor[named]{ACMDarkBlue}{cmyk}{1,0.58,0,0.21}

\hypersetup{colorlinks,
  linkcolor=ACMDarkBlue,
  citecolor=ACMPurple,
  urlcolor=ACMDarkBlue,
  filecolor=ACMDarkBlue}

\definecolor{ckeyword}{HTML}{7F0055}
\definecolor{ccomment}{HTML}{3F7F5F}
\definecolor{cnumber}{HTML}{2A0099}
\definecolor{pblue}{rgb}{0.13,0.13,1}
\definecolor{pgreen}{rgb}{0,0.5,0}
\definecolor{pred}{rgb}{0.9,0,0}
\definecolor{pgrey}{rgb}{0.46,0.45,0.48}

\definecolor{shadecolor}{gray}{1.00}
\definecolor{ddarkgray}{gray}{0.75}
\definecolor{darkgray}{gray}{0.30}
\definecolor{light-gray}{gray}{0.87}

\newcommand{\graybox}[1]{\colorbox{light-gray}{#1}}

\newcommand{\ie}{\emph{i.e.}\xspace}
\newcommand{\aka}{{aka}\xspace}

\newcommand{\eg}{\emph{e.g.}\xspace}

\newcommand{\cf}{\textit{cf.}\xspace}

\newcommand{\set}[1]{\left\{{#1}\right\}}

\newcommand{\Toolname}{{Mallory}}
\newcommand{\jep}{\textsc{Jepsen}\xspace}
\newcommand{\aflfast}{\textsc{AFLFast}\xspace}
\newcommand{\dqlite}{{Dqlite}\xspace}
\newcommand{\braft}{{Braft}\xspace}
\newcommand{\redis}{{Redis}\xspace}
\newcommand{\sqlite}{{SQLite}\xspace}
\newcommand{\tool}{{\sc \Toolname}\xspace}
\newcommand{\toolBold}{{\sc\bfseries \Toolname}\xspace}
\newcommand{\jepsen}{\jep}
\newcommand{\jepsenBold}{{\sc\bfseries Jepsen}\xspace}
\newcommand{\modist}{{\sc MoDist}\xspace}
\newcommand{\flymc}{{\sc FlyMC}\xspace}

\newcommand{\netfilter}{{\sc Netfilter}\xspace}

\newcommand{\afl}{{\sc AFL}\xspace}

\newcommand{\elle}{{\sc Elle}\xspace}
\newcommand{\aflnet}{{\sc AFLNet}\xspace}
\newcommand{\muzz}{{\sc Muzz}\xspace}
\newcommand{\namazu}{{\sc Namazu}\xspace}
\newcommand{\ijon}{{\sc IJON}\xspace}
\newcommand{\sgfuzz}{{\sc SGFuzz}\xspace} 
\newcommand{\stateafl}{{\sc StateAFL}\xspace}

\definecolor{mycolor}{rgb}{0.122, 0.435, 0.698}
\newcommand{\result}[1]{%
\begin{tcolorbox}[colframe=mycolor,boxrule=0.5pt,arc=4pt,
      left=6pt,right=6pt,top=6pt,bottom=6pt,boxsep=0pt,width=\columnwidth]%
      {#1}
\end{tcolorbox}%
}

\newcommand{\ccode}[1]{\mintinline[fontsize=\small]{c}{#1}}
\newcommand{\code}[1]{\ccode{#1}}

%% file: abstract.tex

Grey-box fuzzing is the lightweight approach of choice for finding
bugs in sequential programs. It provides a balance between efficiency
and effectiveness by conducting a biased random search over the domain
of program inputs using a feedback function from observed test
executions.
For distributed system testing, however, the state-of-practice is
represented today by only black-box tools that do not attempt to infer
and exploit any knowledge of the system's past behaviours to guide the
search for bugs.

In this work, we present \tool: the first framework for grey-box
fuzz-testing of distributed systems. Unlike popular black-box
distributed system fuzzers, such as \jep, that search for bugs by
randomly injecting network partitions and node faults or by following
human-defined schedules, \tool is \emph{adaptive}. It exercises a
novel metric to learn how to maximize the number of observed system
behaviors by choosing different sequences of faults, thus increasing
the likelihood of finding new bugs. Our approach relies on \emph{timeline-driven testing}. \tool dynamically constructs Lamport timelines 
of the system behaviour and further abstracts these timelines into
\emph{happens-before summaries}, which serve as a feedback function guiding the fuzz campaign. Subsequently, \tool reactively learns a policy using Q-learning, enabling
it to introduce faults guided by its real-time observation of the summaries.


We have evaluated \tool on a diverse set of widely-used industrial distributed
systems. Compared to the start-of-the-art black-box fuzzer \jep, \tool explores
54.27\% more distinct states within 24 hours while achieving a speed-up of
2.24$\times$. At the same time, \tool finds bugs 1.87$\times$ faster, thereby
finding more bugs within the given time budget. \tool discovered 22 zero-day
bugs (of which 18 were confirmed by developers), including 10 new
vulnerabilities, in rigorously-tested distributed systems such as \braft,
\dqlite and \redis. 6 new CVEs have been assigned.

%% file: intro.tex
\vspace{-5pt}

\section{Introduction}
\label{sec:intro}



Fuzz testing or \emph{fuzzing} is a popular technique for finding
security vulnerabilities in software systems~\cite{fuzz21}. At a high
level, it involves feeding generated inputs into an application with
the goal of finding crashes. Fuzzing can involve a blackbox approach,
where inputs are generated in a purely random fashion, or it can be
guided by knowledge of the program's internal structure (white-box).
The most popular fuzzers are \emph{grey-box}, where the search is guided
by run-time observations of program behaviour, collected, as tests
execute, for artefacts instrumented at compile time.
Thanks to the ease of its deployment and use, grey-box fuzzing is the
state-of-the-practice for automatically discovering bugs in sequential
programs.


A common approach to finding bugs in distributed systems in practice is
\emph{stress-testing}, in which the system is subjected to faults
(\eg, network partitions, node crashes) and its behaviour is checked
against a property-based specification. 
This approach is implemented by tools like \jep~\cite{jepsen}, a
testing framework that is well-known for its effectiveness in finding
consistency violations in distributed databases~\cite{Majumdar2017}.
An alternative to stress-testing is \emph{systematic testing},
commonly known as \emph{software model checking}. In this approach,
the system under test is placed in a deterministic event simulator and
its possible schedules are systematically explored
~\cite{LeesatapornwongsaHJLG14,LukmanKSSKSPTYL19,DragoiEOMN20,ZhouXSNMTABSLRD21,Deligiannis2016}.
The simulator exercises different interleavings of system events by
reordering messages and injecting node and network failures.
Systematic testing is well suited to finding ``deep'' bugs, which require
complex event interleavings to manifest, but is relatively
heavyweight, as it requires integration with the system under test either
in the form of a manually-written pervasive test harness or a system-level
interposition layer. 
While not as effective at finding deep bugs, stress-testing is widely
used due to its low cost of adoption and good effort-payout ratio.


\vspace{3pt}
\paragraph{\emph{Problem statement}}
We make the following observation:
in terms of the ease-of-use/effectiveness trade-off, black-box fuzzing of
sequential programs is similar to stress-testing of distributed
systems, while white-box fuzzing corresponds to software model checking.
However, unlike in the sequential case, there is no grey-box fuzzing
approach for distributed systems.
Our goal is to explore this opportunity by extending \jep with the
ability to \emph{perform observations} at runtime about the behaviour
of the system and to \emph{adapt} its testing strategy based on
feedback derived from those observations.
In doing so, we are not aiming to match the thoroughness of systematic
testing, but to provide a more effective and principled way to conduct
stress-testing while maintaining its ease of use.


\vspace{3pt}
\paragraph{\emph{Challenges}}
In the last decade, developing a grey-box fuzzer for sequential programs
has become more streamlined, due to fuzzers like AFL~\cite{afl}. 
In short, AFL works by generating and mutating inputs to a program
being tested, aiming to trigger crashes or other unexpected behavior.
It uses a feedback-driven approach, keeping track of inputs
that cause the program to take \emph{new code paths} and prioritizing
mutations that are likely to explore these paths further.
Attempting to adapt the greybox approach to \jep-style distributed system
testing leads us to three questions:
\begin{enumerate}[label=\textbf{Q\arabic*},topsep=2pt,leftmargin=17pt]
\item \label{q1} What is the space of inputs to a distributed system that
could be explored adaptively?
\item \label{q2} What observations are relevant for a distributed
  system and how should they be represented?
\item \label{q3} How can one obtain feedback from the observations?
\end{enumerate}
Question~\ref{q1} is already answered by \jep: the role
of~``inputs'' for distributed systems is played by \emph{schedules},
that can be manipulated by injecting faults.
Even though \jep can control the fault injection, in the absence of a
good feedback function, it (a) requires human-written generators to
explore the domain of schedules if something more than random fault
injection is required~\cite{Alvaro2017} and (b) repeatedly explores
equivalent schedules.

To answer~\ref{q2} we recall perhaps the most popular graphical formalism to
represent interactions between nodes in distributed systems: so-called
\emph{Lamport diagrams} (\aka \emph{timelines}), \ie, graphs showing relative
positions of system events as well as causality relations between
them~\cite{Lamport78,fidge1988timestamps,mattern1989virtual}.
Such diagrams have been used in the past for visualizing executions in
distributed systems~\cite{BeschastnikhLXW20}.
Our discovery is that they also can be used as distributed analogues
of ``new code paths'' from sequential grey-box fuzzing. In other
words, being able to observe and record new shapes of Lamport diagrams
is an insight that brings AFL-style fuzzing to a distributed world.

To make our approach practical, we also need to address~\ref{q3}.
The problem with using observed Lamport diagrams to construct a
feedback function is that in practice no two different runs of a
distributed system will produce the \emph{same} timeline.
That is, such new observations will \emph{always} produce new feedback,
even though in practice many runs are going to be equivalent for the
sake of testing purposes---something we need to take into
account.\footnote{A sequential analogue of concrete distributed
  timelines would be a trace of \emph{all} memory operations---too
  precise to recognise equivalent executions.}
As a solution, we present a methodology for extracting feedback from
dynamically observed timelines by \emph{abstracting} them into concise
\emph{happens-before summaries}, which provide the desired trade-off
between the feedback function's precision and effectiveness.

\vspace{3pt}
\paragraph{\emph{Contributions}}
The solutions to~\ref{q1}--\ref{q3} provide a versatile conceptual
framework for grey-box fuzzing of distributed systems.
%
%
Building on these insights, we present our main practical innovation:
\tool, the first grey-box fuzzer for distributed systems. 
Unlike the black-box testing approach of \jep that requires
human-written schedule generators, \tool reactively \emph{learns} them
by (a) observing the behaviour of the system under test as it executes
and (b) rewarding actions that uncover new behaviour. 
Below, we detail the design, implementation, and evaluation of \tool.
\begin{itemize}
\item \emph{Timeline-driven testing}, a novel fuzzing approach suited for
distributed systems: It is based on dynamically constructing Lamport diagrams
(timelines) of the system under test as it runs, and further abstracts the timelines
into happens-before summaries. It is used to define a \emph{feedback function} guiding grey-box fuzzing.
\item \emph{Reactive fuzzing}, a reactive method for making optimal decisions 
to achieve maximized behaviour diversity: It reactively learns a policy using Q-learning
to decide what actions to take in observed states, to incrementally construct a schedule.

%
\item \emph{End-to-end implementation} of \tool, a fuzzing framework for
distributed systems: \tool extends the widely used \jep framework---\tool can
be seen as an adaptive generator of schedules for \jep tests. The tool is publicly available
at 

\begin{center}
    \url{https://github.com/dsfuzz/mallory}
\end{center}
\item \emph{Comprehensive evaluation} of \tool on several widely-used
industrial distributed system implementations. In our experiments, \tool covers
54.27\% more distinct states within 24 hours and achieves the same state
coverage about 2.24$\times$ faster than \jep. In terms of reproducing existing
bugs, \tool speeds up the bug finding by 1.87$\times$ and finds 5 more bugs
compared to \jep. Moreover, in rigorously-tested distributed systems, \tool
found 22 previously unknown bugs, including 10 new security vulnerabilities and
6 newly assigned CVEs. Out of these 22 bugs, 18 bugs have been confirmed by
their respective developers. In our experiments, \jep could only detect 4 of
these bugs.

\end{itemize}

%% file: overview.tex
\section{Overview}
\label{sec:overview}

In this section, we illustrate the workflow of our technique for
adaptively detecting anomalies in distributed systems.

\subsection{Bugs in Distributed Systems}
\label{sec:raft}

\begin{figure}[t]
\centering
\begin{minted}[fontsize=\footnotesize,xleftmargin=0pt,]{c}
145    int membershipRollback(struct raft *r){
146      ...
158      // Fetch the last committed configuration entry 
159      entry = logGet(&r->log, r->config_index);
160      assert(entry != NULL);
176    }

986    static int deleteConflictingEntries(){
987      ...
1007     // Possibly discard uncommitted config changes
1008     if (uncommitted_config_index >= entry_index){
1009       rv = membershipRollback(r); 
1010     }
1042   }
\end{minted}
\caption{Simplified \dqlite code for membership rollback.}
\label{fig:rollback}
\end{figure}

As a motivating example, let us consider a known bug in the implementation of
the Raft consensus protocol~\cite{OngaroO14} used by \dqlite, a widely-used distributed
version of \sqlite developed by Canonical.\footnote{Available at
\url{https://dqlite.io}; 3.4k stars on GitHub at the time of writing.}

The purpose of using a \emph{consensus protocol} in a distributed system is to
ensure the system maintains a consistent and reliable state even in the
presence of faults. In Raft, one of the most widely used consensus protocols, a
single leader accepts client requests and replicates them to all nodes that
persist them as log \emph{entries}.
Conflicting entries in a Raft cluster can appear when different nodes receive
different log entries during a network partition.
Over time, the number of replicated entries might grow very large, which, in
turn, might cause issues if certain nodes need to be brought up-to-date after
having experienced a temporary downtime.
To address this issue, Raft periodically takes a \emph{snapshot} of the current
system state, discarding old log entries whose outcome is reflected in the
snapshot.
Additionally, nodes can be removed from the cluster or join it,
thus changing the configuration of the system as it runs. 
When a configuration change is initiated, the current leader
replicates a \emph{configuration change entry} to all the nodes in the
cluster.
A new configuration becomes permanent once it has been agreed upon and
committed by a majority of the nodes, yet a server starts using it as soon as
the configuration entry is added to its log, even before it is
committed~\cite[\S{6}]{OngaroO14}---a fact that is important for our example.
If there is a failure during the process of agreeing on a new configuration,
such as a network partition, the new configuration may not be fully replicated
to the majority, in which case the leader node will attempt to perform a
\emph{membership rollback} by adopting the last committed configuration entry
from its log.

The bug in question occurs during a membership rollback happening
\emph{immediately after} performing a snapshot operation, leading to
a failure to restore the last committed
configuration~\cite{membership}.
\autoref{fig:rollback} shows the affected fragment of the actual
implementation in \dqlite, which deals with removing conflicting
entries during node recovery.
In case there is an uncommitted configuration entry among the
conflicting entries to remove, a \dqlite server has to first roll back
to the previously committed membership configuration via
\code{membershipRollback} (line~\code{1009}).
When this happens after a snapshot operation, which has removed the last
committed configuration entry, the assertion on line~\code{160} gets
violated.

\begin{figure}[!t]
\setlength{\belowcaptionskip}{-15pt}
\setlength{\abovecaptionskip}{2pt}
  \centering
  \includegraphics[page=1, trim= 0in 0in 0in 0in, clip, scale=0.5]{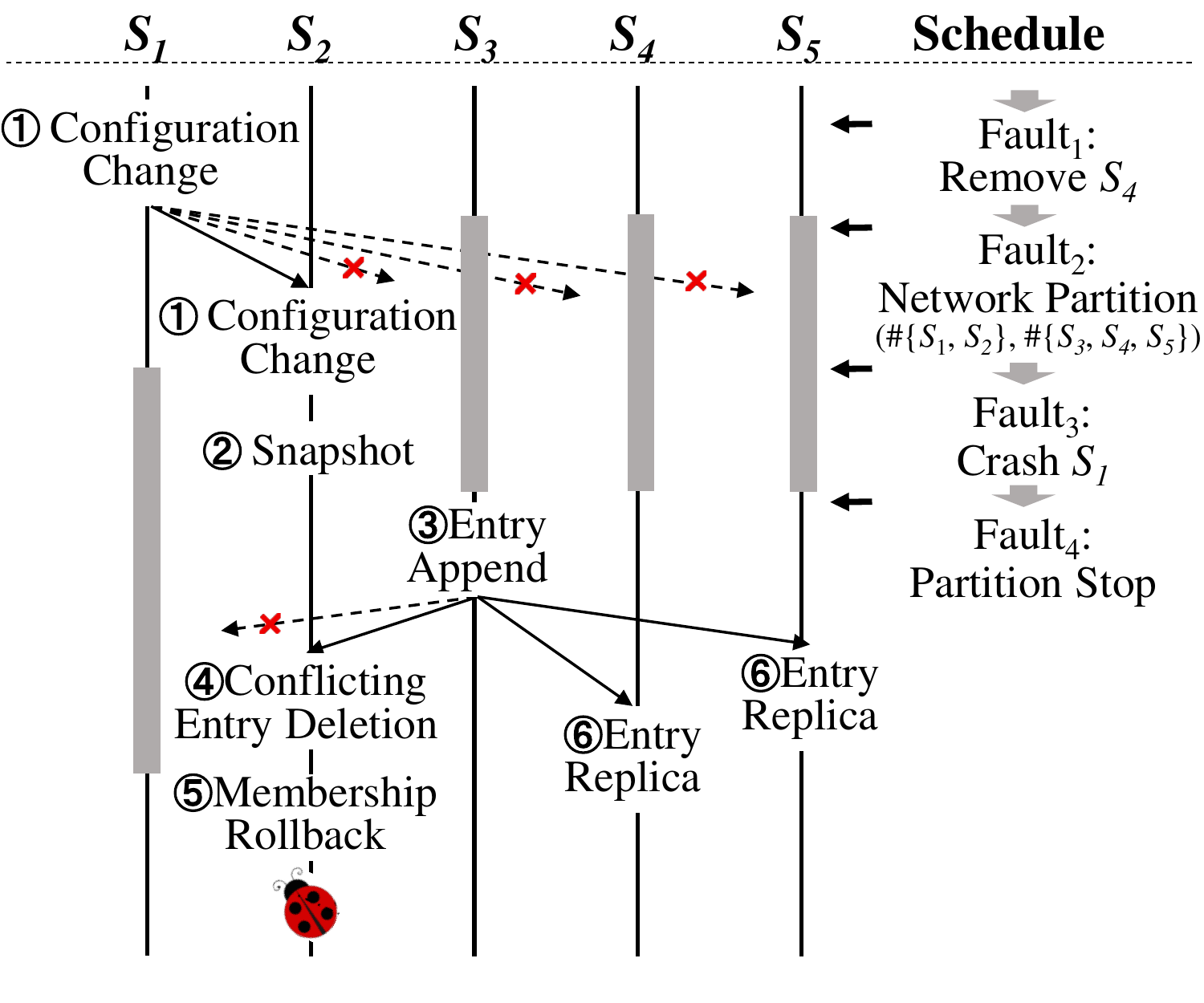}
  \caption{A timeline of the \dqlite membership rollback bug. Gray
    vertical rectangles correspond to node downtimes.}
  \label{fig:timeline}
\end{figure}


%
To show how this rather subtle bug can be triggered in a real-world
environment, consider a run of a \dqlite cluster depicted
in~\autoref{fig:timeline}.
The initial cluster comprises five servers $S_1$--$S_5$, with  $S_1$
assumed to be a leader. 
Server $S_4$ requests (to the leader $S_1$) to be removed from the cluster.
Upon receiving this request, leader $S_1$ appends the configuration
change entry into its log (\ding{172}) and attempts to replicate it
to all other members, but only succeeds to do so for $S_2$
(\ding{172}), failing to reach $S_3$, $S_4$, and $S_5$ due to
a sudden partition in the network. 
At the same time, the number of log entries at the server $S_2$
reaches a threshold value, prompting it to take a snapshot
(\ding{173}), while \emph{already using} the latest configuration
(which has not been agreed upon by the majority), thus, discarding the
old configuration entry from the snapshot.
To make things worse, $S_1$ crashes at the same time.
Although the network of $S_3$, $S_4$, and $S_5$ recovers soon after,
the cluster has already lost its leader. 
At some point, a new leader election is initiated (details omitted),
with $S_3$ eventually becoming the leader and attempting to
synchronise logs across the nodes (\ding{174}).
Prompted to do so, $S_2$ detects a conflicting entry in its log (\ie,
the uncommitted configuration change (\ding{172})) and deletes it
(\ding{175}).
It then attempts to retrieve the last committed configuration entry to
roll back the membership (\ding{176}), which is long gone due to the
prior snapshotting (\ding{173}), triggering the assertion violation at
line~\code{160} of \autoref{fig:rollback}.

\subsection{Fuzzing Distributed Systems via \jep}

As demonstrated by the example, identifying a bug in a distributed
system in some cases boils down to constructing the right sequence of
\emph{faults}, such as network partitions and node removals, resulting
in an execution that leads to an inconsistent state and, subsequently,
to the violation of a code-level assertion or of an externally
observable notion of consistency (\eg,
linearisability~\cite{HerlihyW90}).
The state-of-the-art fuzzing tool \jep provides a means to randomly
generate sequences of faults with the goal of discovering such bugs.

\autoref{algo:jeppa} provides a high-level overview of the workings of
\jep (let us ignore the grayed fragments for now).
\jep requires a lightweight harness for the system under test to
define how to start and stop it, to enact faults, introduce client
requests, and collect logs (line~1).
For brevity, \autoref{algo:jeppa} does not show the set-up of the SUT at the
beginning of each test or the introduction of client requests, which the SUT is
constantly subjected to by client processes.
Importantly, \jep allows the user to define testing \emph{policies} (\aka
``generators'') responsible for introducing specific types of external inputs or
faults (line~2).
%
%
%
The main fuzzing loop of \jep is shown in lines 3--16 of the
algorithm.

\begin{algorithm}[t]
{\small{
  \DontPrintSemicolon
  \LinesNumbered
  \SetNoFillComment
  \KwIn{$P_0$: system under test (SUT)}
  \KwIn{$\it Nem$, $\it Faults$: a nemesis and the faults it can enact}
  \KwIn{$\it Oracles$: a set of test oracles for bug detection}
  \KwIn{$S$: number of steps in each schedule}
  \KwIn{$T$: total time budget for testing}
  \KwOut{$\it Bugs$: a set of bugs detected}
  $P_f$ $\gets$ {\graybox{instrumentSystem}}($P_0$)\;
  $\it Policy$ $\gets$ \{\graybox{$\it initState$}, $\it Faults$\}\;

  \Repeat{\textup{time budget $T$ exhausts}}{
    \graybox{$\it curState$ $\gets$ $\it initState$} \;
    \Repeat {\textup{maximum steps $S$ reached}} {
        $\mathit{fault}$ $\gets$ $\it Policy$.getNextFault (\graybox{$\it curState$})\;
        $\mathit{Nem}$.enactFault($\it fault$)\;
        $\it events$ $\gets$ observeSystemUnderTest ($P_f$)\;
        \graybox{$\it timeline$ $\gets$ constructTimeline ($\it events$)}\;
        \graybox{$\it nextState$ $\gets$ abstractTimeline ($\it timeline$)}\;
        \graybox{$\it rwd$ $\gets$ calculateReward ($\it curState$, $\it fault$, $\it nextState$)}\;
        \graybox{$\it Policy$ $\gets$ learn ($\it Policy$, $\it curState$, $\it fault$, $\it rwd$)}\;
        \graybox{$\it curState$ $\gets$ $\it nextState$}\;
    }
    
    resetSystemUnderTest ($P_f$)\;
  }
  $\mathit{Bugs}$ $\gets$ $\mathit{Oracles}$.identifyBugs ($\it events$)\;
  
\caption{Fuzzing with {\jep} and \graybox{\tool}}
\label{algo:jeppa} 
}}

\end{algorithm}

During each run of the outer loop, the framework generates a system-specific
external input or fault (line~6) via a \emph{policy}, and enacts it using a
\emph{nemesis}---a special process, not bound to any particular node, capable
of introducing faults. Such inputs may, for example, be the decision to remove
a node from the system, as, \eg, is done by node $S_4$ in our running
example.
As the system is executing, the framework records its observations (line~8) for
future analysis to detect the presence of bugs or specification violations
(line~17).
This process continues until the time budget $T$ is exhausted (line~16).
The test run is segmented into schedules of~$S$ steps each, after which
the system is reset (line~15). 
%


Getting back to our example, we can see that the membership rollback
bug can be exposed by the scheduled sequence of inputs/faults that
first initiates the removal of~$S_4$ from the cluster and then creates
a network partition $(\#\!\set{S_1, S_2}$, $\#\!\set{S_3, S_4, S_5})$,
followed by node a crash of~$S_1$. Randomly generating this particular
sequence of faults via \jepsen, while possible, is somewhat unlikely.
The reason is: before coming across this schedule, \jepsen may try many others,
each making very little difference to the system's observable behaviour, \eg,
by randomly crashing a number of nodes. In our experiments, \jepsen failed to
detect this membership rollback bug (\ie, Dqlite-323 in
\autoref{sec:reproducing}) within 24 hours.

However, with just a little insight into the system, one can
conjecture that enacting a partition right after a configuration change
leads to novel system states more often than, \eg, performing another
configuration change, thus, increasing the likelihood of witnessing a
new, potentially bug-exposing, behaviour.
Our goal is to retrofit \jepsen so it could derive these insights at
run time and adapt the policies accordingly.


\subsection{Learning Fault Schedules from Observations}

The high-level idea behind \tool, our fuzzing framework, is to enhance \jepsen
with the ability to \emph{learn} what kinds of faults and fault sequences are
most likely going to result in previously unseen system behaviours.\
To achieve that, we augment the baseline logic of \autoref{algo:jeppa}
by incorporating the \graybox{grayed} components that keep track of
the observations made during the system runs.
The first change is to add instrumentation to the system under test (line~1) to
record significant \emph{events} (\eg, taking snapshots or performing
membership rollbacks in~\autoref{fig:timeline}) during the execution,
additional to those \jep already records, \ie, client requests, and responses.
More interestingly, the fault injection policy is now determined
not just by the kinds of faults and inputs that can be enacted, but by
the latest \emph{abstract state} of the system, whose nature will be
explained in a bit and that is taken to be some default
\emph{initState} at the start of the fuzzing campaign (line~2).

The main addition consists of lines~9-13 of the algorithm.
Now, while running the system, the fuzzer collects sequences of events
recorded by the instrumented nodes, as well as message-passing
interactions between them; the exact nature of events and how they are
collected will be described in~\autoref{sec:observe}.
The information about the recorded events and their relative ordering
is then used to construct a (Lamport-style) \emph{timeline} and
subsequently \emph{summarised} to obtain the new abstract state
\emph{nextState} (lines~9-10)---the design of these two procedures,
detailed in~\autoref{sec:timelines}, is the central technical
contribution of our work.
The newly summarised abstract state is used to calculate the reward
\emph{rwd} by estimating how dissimilar it is compared to abstract
states observed in the past (line~11).
Finally, the reward is used to dynamically update the policy, after
which the loop iteration repeats with the updated abstract state (lines~12-13).

Postponing until~\autoref{sec:design} the technicalities of computing
abstract states, calculating rewards, and updating the policy, let us
discuss how the introduced changes might increase the likelihood of
discovering the bug-inducing system behaviour
from~\autoref{fig:timeline}.
We now pay attention to the six kinds of events
(\ding{172}-\ding{177}) that can be recorded in the system, as well as
their relative happens-before ordering is computed across multiple
nodes.
Consider a fault injection policy that introduces a sequence of node
removals (such as $\text{Fault}_1$). After triggering several
configuration changes (\ie, event~\ding{172}), such a policy will not
introduce many new behaviours in a long run, which will prompt our
adaptive fuzzer to prefer other faults, \eg, network partitions.
By iterating this process, observing new behaviours (\ie, different
event sequences) in the form of novel abstract states and
de-prioritising policies that have not generated new behaviours, the
fuzzer will eventually discover a sequence of faults leading to the
membership rollback bug.

It is important to note that the fact that a particular policy has not produced
a new abstract state (\ie, a new observable behaviour) in a particular run does
not necessarily mean that it needs to be discarded for good. Due to the nature
of the applications under test, \tool, similarly to \jepsen, does not provide a
fully deterministic way to inject faults, hence some behaviours might depend on
the absolute timing of faults.
This is taken into account by \tool's learning
(\cf~\autoref{sec:learn}), which leaves a possibility for such a
policy to be picked again in the future, albeit, with a lower
probability.

In our experiments, due to \tool's adaptive learning, the membership rollback
bug was discovered in 8.68 hours (\jep failed to discover it in 24 hours). In
the following, we give a detailed description of \tool's design
(\autoref{sec:design}) and provide thorough empirical evidence of its
effectiveness and efficiency for discovering non-trivial bugs in distributed
systems (\autoref{sec:evaluation}).


%% file: design.tex
\section{The \toolBold Framework}
\label{sec:design}

At its core, \tool implements an adaptive \emph{observe-orient-decide-act}
(OODA) loop:

\begin{itemize} 
    
\item\emph{Observe}---observe each node's internal behaviour and intercept all
network communication between nodes;

\item\emph{Orient}---construct a global Lamport timeline of the system's
behaviour to obtain a bird's eye view of the execution, and abstract the
timeline into a manageable representation, called a happens-before summary,
used to understand the current state of the system and to determine the
effectiveness of previous actions;

\item\emph{Decide}---choose a fault to inject based on the current observed
summary and the past execution history;

\item\emph{Act}---inject the fault and repeat the loop.
\end{itemize} 

\begin{figure*}[ht]
\setlength{\belowcaptionskip}{-10pt}
\setlength{\abovecaptionskip}{5pt}
    \centering
    \includegraphics[width=0.95\textwidth]{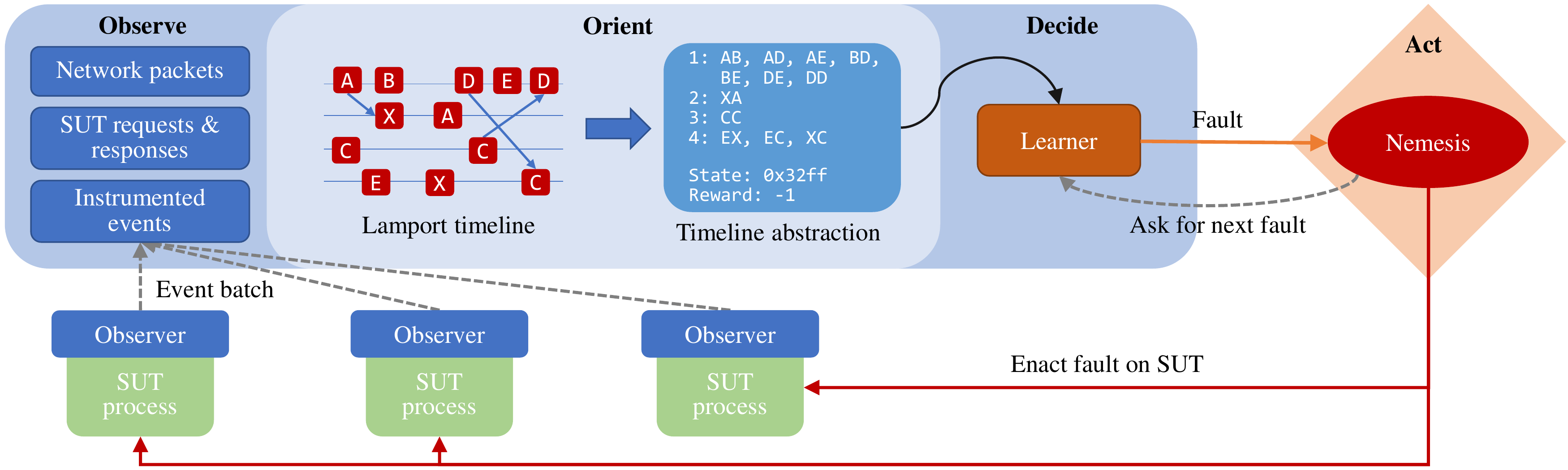}
        
    \caption{The central observe-orient-decide-act loop in \toolBold. A centralised
    \emph{mediator} collects events from \emph{observers} distributed at the
    nodes in the SUT, and drives the test execution. Faults decided by \toolBold
    are enacted by \jep. }
    
    \label{fig:ooda-loop}
\end{figure*}

%
Unlike white-box fuzzers, which rely on encapsulating the system under
test in an event simulator~\cite{DragoiEOMN20,OzkanMO19}, \tool operates
on the actual system in its normal distributed environment---a firm
requirement to minimise the friction (\ie, adoption effort).
In particular, \tool does not have the luxury of being able to
``pause'' the system and observe its state before deciding what
actions to take, as it operates in real-time, in a reactive manner.
This means \tool itself is a distributed system, which
complicates its implementation slightly. Nonetheless, its architecture
is designed to hide this as much as possible from users, as will
become apparent.

To explain \tool's design, we will walk through an entire
observe-orient-decide-act loop, step by step, gradually introducing
its architectural components.


\subsection{Observing the System Under Test}
\label{sec:observe}

\tool's first task is to observe the system under test~(SUT). Broadly, there
are three types of observations that we can make: (1) network
observations, which capture communication between nodes in the SUT
(\eg, a packet was sent from node A to node B and received by node B),
(2) external observations, which capture the input-output behaviour of
the system (\eg, requests and responses for a database), and (3)
internal observations, which capture a node's internal behaviour (\eg,
a function was executed, a conditional branch was taken, an error
message was logged). In the following, we use ``observation'' and
``event'' interchangeably.

Events happen on a particular node at a particular time. However, as
is well known, in a distributed system there is no globally shared
notion of time. 
We postpone the explanation of how \tool constructs a global
timeline without assuming precise clock synchronisation and without
tagging messages with vector clocks. 
For now, it suffices to say that each event carries a node identifier
and a \emph{monotonic timestamp} returned by the node's system clock.

Below, we outline how \tool observes the defined above types of events
capturing the patterns of communication~(\autoref{sec:event1}),
externally observable input/output~(\autoref{sec:event2}), and
internal behaviour of the nodes~(\autoref{sec:event3}).

\subsubsection{Packet Interception.}
\label{sec:event1}

To keep our framework lightweight and require as little modification
of the system under test as possible, we capture TCP and UDP packets
at the IP network layer using Linux's firewall infrastructure, rather
than require users to instrument the application layer to identify
protocol-level messages.


By necessity, \tool's architecture is distributed, matching the
structure of the SUT. As shown in \autoref{fig:ooda-loop}, \tool
consists of a number of \emph{observer} processes, one at each node,
that observe local events (bottom half of the figure), and a central
\emph{mediator} process that collates information from all observers
and coordinates the execution of the test (large blue rectangle in the
top half).
At every node, the observer, which the \jep test harness starts before
the system under test, installs a \netfilter firewall queue that
intercepts all IP packets sent to or from the node. During the test,
the kernel copies packets to the observer process in user space, where
each packet is assigned a monotonic timestamp, recorded, and then
emitted unchanged.


\paragraph{\emph{Mediator interception}} 
Observers collect packet events in batches and forward those to the
mediator periodically, by default every 100ms. Rather than include the
entire packet in the batch, which would entail trebling network
traffic, observers only record and send to the mediator a 64-bit
packet identifier obtained from the source and destination IP
addresses and ports and from the IP and UDP or TCP headers'
identifiers, respectively. Yet we do want the mediator to have access
to the packet contents: for instance, the content of messages might
determine what is the best fault to introduce. To achieve this, we set
up the test environment that the SUT executes in such that \emph{all
  packets pass through the node running the mediator}. Concretely, we
place each node on its own separate (virtual) Ethernet LAN, with the
mediator acting as the gateway for all the LANs. The \tool mediator
acts as a \emph{man-in-the-middle} for all packets in the SUT. 
It can then reconstruct the identifier for each packet, and
cross-reference it with the batches received from the sending and
receiving nodes' observers to determine the respective timestamps.
The mediator, unlike the observer, which is passive, is active and implements a
full user-space firewall using \netfilter. It can delay and drop packets when
instructed to do so by the decide step of the OODA loop. 

Using this infrastructure, the mediator builds up a complete picture of the
system under test's communication.

\subsubsection{Requests and Responses.}
\label{sec:event2} 


The observations about input/output of nodes are made at the
application layer.

We built \tool on top of \jep, and reuse \jep's infrastructure to
define the test harness that: (a)~sets up and starts the SUT,
(b)~defines and executes a workload (a sequence of client requests to
the SUT), (c)~enacts faults (\eg, crashing a node), and (d)~checks the
validity of the SUT's response to the workload. 
\jep already captures requests and responses for validity checking (\eg, for
linearisability), and we hook into this existing code, attach monotonic
timestamps to events, and relay them to the mediator reactively.
%






\subsubsection{Code Instrumentation}
\label{sec:event3}

The final kind of observation we make is at the code level. We want \tool to be
able to peek into the internal workings of the SUT, beyond what is observable
to clients of the system or to eavesdroppers on the network. For this, we reuse
the compile-time instrumentation infrastructure used by greybox fuzzers (\eg, \afl) for
sequential programs. Like those fuzzers, \tool adds
\emph{instrumentation code} to the SUT to capture and expose runtime
information about the program's execution. The key question is: what about the
execution should we capture?


In our early experiments, we used the notion of \emph{edge coverage}, the type
of instrumentation that has become standard for fuzzing sequential programs due
to its empirically-observed effectiveness. It maintains a global bitmap of
code edges, and increments an approximate counter for each edge
that is traversed during program execution. At the end of the execution,
the bitmap serves as a summary of ``what the program did,'' and is used by the
fuzzer to assign energy and mutate its input during subsequent runs. This is a great metric for
certain kinds of programs, \eg, command-line utilities and file-parsing
libraries, but---as we quickly discovered---not particularly meaningful for
distributed systems. The goal in fuzzing sequential programs is to generate
inputs that go ``deep'' into a program and explore all ``cases'' (\ie,
conditional branches). For such programs, the thoroughness of exploration is
naturally defined in terms of code coverage. But this is not the case at all
for distributed systems. Distributed systems tend to be implemented as reactive
event loops and run almost the same code for every request, with minor
variations. Code coverage metrics tend to saturate very quickly when testing such
systems.

A natural behavioural metric for distributed systems, which we came to adopt,
is that of the \emph{event trace}. Executions in a distributed system are
distinguished not so much by which events happen, but by the order in which
they happen. Moreover, as has been empirically observed, what tends to uncover
bugs are specific subsequences of events, \eg, $A$ before $B$ before $C$, with
potentially many events between them~\cite{YuanLZRZZJS14}. The disadvantage of
event traces compared to code coverage is that the former can become very large
and expensive to store and operate on, especially if every basic block is instrumented. To
alleviate this issue, for now we require from the user a small amount of manual
annotation of the SUT's code, in the form of \ccode{//INSTRUMENT_FUNC}
and \ccode{//INSTRUMENT_BLOCK} comments, to indicate which 
basic blocks and functions
are ``interesting'' and should be tracked by \tool. 

Our instrumentation creates a
POSIX shared memory object accessible by the observer process, and stores in it
a fixed-size global array of events along with an atomic index.
We implement a LLVM pass that assigns a unique ID to every annotated basic
block and function in the SUT, and inserts the hooking code at the start of the
block or function. During program execution, this code gets a monotonic timestamp and records the event
in the global array at a fresh position. 
The observer process periodically reads the
shared memory object, copies the trace, and resets the counter; it also
includes the trace in the periodic batch it sends to the mediator.
%



\subsection{Making Sense of Observations}
\label{sec:timelines}

For the second phase of the loop, \tool needs to make sense of the events it
received from observers. The goal of this phase is to transform the ``raw''
event data into a form more amenable to analysis and decision-making.

It is at this stage that our key conceptual contributions of
\emph{timeline-driven testing} and \emph{timeline abstraction} come into play.
At the core of \tool's OODA loop lie its \emph{view} of the world, a
dynamically constructed Lamport timeline of events in the SUT, and its
\emph{model} of the world, a user-defined abstraction of the timeline. \tool
first builds a birds-eye view of the SUT's execution by constructing a global
timeline, then makes sense of the timeline by abstracting it into a summary
consisting of its ``essential'' parts, which is used to judge the effectiveness
of previous actions and to decide which faults to introduce next.

\subsubsection{Building the Timeline.} 

\begin{algorithm}[t]
{\small{
  \DontPrintSemicolon
  \LinesNumbered
  \SetNoFillComment
  \KwIn{Map from node IDs to ordered sets of events}
  \KwOut{Prefix-closed causal timeline as a graph}
  $(\var{prefixRanges}, \var{extensionRanges}) \gets \fun{readyRanges}()$\;
  $\var{links} \gets \fun{trackLinkSources(\var{extensionRanges})}$\;
  $\var{rw} \gets \fun{map}(\fun{first}, \var{prefixRanges})$ \tcc{resumeWith}
  $\var{ln} \gets \fun{map}(\fun{last}, \var{prefixRanges})$  \tcc{lastNeeded}
  \While{\textup{$\exists n.~\var{rw}[n] \neq \var{ln}[n]$}}{
    \For{\textup{$n \in \var{nodes}$}}{
      \While{\textup{$\var{rw}[n] \neq \var{ln}[n]$}}{
        $\var{range} \gets \var{rw}[n] \isep \var{ln}[n]$\;
        \For{\textup{$\var{ev} \in \var{range}$}}{
          \If{\textup{$\var{ev} \text{ is the target of } \var{src} \in \var{links}$}}{
            $\fun{attachLinkTarget}(\var{ev}, \var{links})$\;
            $\var{ln}[\fun{node}(\text{src})]\! \gets \! \fun{max}(\var{src}, \var{ln}[\fun{node}(\text{src})])$\;
          }
          $\var{rw}[n] \gets \var{ev}$\\
        }
      }
    }
  }
  Add all events and links up to $\var{ln}[n]$ to the graph.\;
}}  
  \caption{Build prefix-closed Lamport timeline}
  \label{algo:timeline-construction} 
\end{algorithm}

As the system is executing, the \tool mediator receives batches of events from
all the observers and adds them into a single global timeline. Every event is
associated with a particular node in the SUT and has an attached monotonic
timestamp from that node's system clock. However, events \emph{do not} have
causal timestamps (\eg, vector clocks) that encode the causal relationship
between events at different nodes. In other words, the mediator at this stage
has a timeline with events, but no causal arrows between events on different
nodes. It must use its complete knowledge of the system's communication to
reconstruct the causal ordering of observed events.

Importantly, the timeline that is passed on to the next stage to be abstracted
must be \emph{prefix-closed}, in the sense that if an event $e$ is included,
all its causal predecessors must also be included. Due to the distributed
nature of the system, there are some complications in creating prefix-closed
timelines: (a) batches of events from a single observer might arrive out of
order and (b) receipt events may arrive before their respective send events (in
other words, the target of a causal arrow may arrive before the source). The
first issue is straightforward to solve: give a sequence number to each batch,
and have the mediator ensure batches are added to the timeline in sequential
order. This is sufficient to guarantee that each node's local timeline is
prefix-closed. The second issue is more challenging due to the recursive nature
of prefix-closedness across nodes. Recall that the mediator now has a timeline
with events, but no causal arrows. The question is: how can we know that we
have received all the causal predecessors of a given event, \ie, that an event
belongs to the prefix-closed portion of the timeline?

There are multiple ways to solve this problem. The approach we choose exploits a
well-known property of causality: real-time order is an over-approximation of
causal order, \ie, if event $B$ happens in real-time after event $A$, then $B$
cannot causally influence $A$. In other words, all events $A$ that causally
influence an event $B$ must be before $B$ in real-time order. This means that
we only need to look for an event's causal predecessors in the timeline up to
the point where the event's real-time timestamp is first exceeded. The issue is
that (c) events are tagged with monotonic, not real-time timestamps, and (d)
nodes in any case do not have synchronised real-time clocks. We address (c) by
requiring each observer to submit upon start-up both a monotonic timestamp and
a real-time timestamp obtained at roughly the same time. This lets us
approximately convert monotonic timestamps at a single node into real-time
timestamps for that node. 
The issue (d) of clocks not being
synchronised across nodes still remains, however. The solution: we introduce a
bound on the maximum clock skew between nodes (we set it to a conservative
100ms), and use this to limit ``how far in the future'' we look for causal
predecessors of an event $e$ on other nodes' timelines, relative to $e$'s
real-time timestamp.

\autoref{algo:timeline-construction} shows our timeline construction procedure.
It takes as input a map from node IDs to the ordered set of events at that
node, and constructs a prefix-closed causal timeline (\aka a
\emph{consistent cut}~\cite{mattern1989virtual}) in the form of
a graph---events are nodes and causal arrows are edges. The algorithm proceeds
in four stages. First, on line 1, we identify for each node the range of events
that have not yet been processed and whose causal predecessors must all exist
in the input (\texttt{prefixRanges}). This is obtained by subtracting the
maximum clock skew ($cs$) from the last real-time timestamp ($ts$) up to which
\emph{all} nodes have submitted events. We are guaranteed to have all the
causal predecessors of events in this range---these will be found in the range
up to $ts + cs$ on other nodes (\texttt{extensionRanges}). Second, on line 2,
we traverse the extension ranges, identify all events that are the source of
inter-node causal arrows (\ie, packet sends, which, recall, have 64-bit
identifiers), and keep track of these causal arrows, identified by the source's
unique ID (\texttt{links}). The third stage (lines~5--13) is a fixpoint
computation that keeps track of where the prefix-closed portion of the timeline
ends (\texttt{lastNeeded}) for each node and ``fills up'' the timeline by
traversing it (lines~8--9) and ``requiring'' that causal predecessors of
encountered events also be in the timeline (lines~11--12). The fixpoint
computation terminates (line 5) when there are no gaps for any nodes, \ie, the
timeline is prefix-closed up to $\var{ln}[n]$ for all nodes $n$. The fourth and final
stage (line 14) is to construct a graph from the prefix-closed portion of the
timeline and pass it to the abstraction phase. (The graph is prefix-closed, \ie,
all receive have a matching send, but not necessarily suffix-closed, \ie, there
may be sends with no matching receive. The causal arrows for these point to a
special node at infinity, and will point to the correct receive when it is
encountered later.)

\subsubsection{Timeline Abstraction}

\begin{figure}[t]
\setlength{\belowcaptionskip}{-10pt}
\setlength{\abovecaptionskip}{5pt}

\begin{minted}[fontsize=\small, linenos,numbersep=12pt,xleftmargin=20pt]{rust}
pub trait TimelineAbstraction {
  fn update(&mut self, ev: &Event);
  fn merge(&mut self, this_ev: &Event,
            other: &Self, other_ev: &Event); 
}
\end{minted}

\caption{Rust interface for defining timeline abstractions. Abstractions are
built incrementally by iterating over the causal structure, storing ``what
matters'' along the way.}

\label{fig:summary-trait}
\end{figure}

We now have a Lamport timeline describing the system's behaviour, obtained in
almost real-time, and want to use it to drive our testing campaign. More
concretely, we want our decisions to adapt the system's behaviour and drive
the execution towards new behaviours. But what counts as novel behaviours? A
naive approach is to use the timeline itself as feedback for our decision: 
we want to see timelines different from what we have seen before. This
does not work because the timeline is a low-level representation of
the system's behaviour; all observed timelines are unique, even discounting
event timestamps and packet contents. Clearly, to be able to operate
effectively with timelines, we need in some fashion to \emph{abstract} them
into ``what really matters''. Eliminating timestamps and packets is a form of
abstraction, but it is not enough: we need something a bit more clever.

Importantly, we do not want to bake into the tool any particular notion of
``what really matters''. (We do provide sensible defaults.) Instead, we want to
allow users of \tool to specify what is important for their particular systems
and testing needs. Moreover, we want an intuitive interface for specifying
this, one that any distributed system engineer can understand and use. To this
end, we introduce our novel notion of \emph{timeline abstraction}, inspired by
vector clocks. Our insight is that the principle used to define causal
timestamps, namely that of \emph{accumulating causal information} via
\mintinline{rust}|update| (\aka \emph{event} or \emph{copy}) and
\mintinline{rust}|merge| (\aka \emph{join})
operations~\cite{fidge1988timestamps, mattern1989virtual}, can be generalised
to arbitrary abstractions of causal diagrams. 
The interface for defining such abstractions is shown in \autoref{fig:summary-trait}. 
Intuitively, a timeline abstraction is attached to an event and represents or \emph{abstracts} the
whole causal history up to (and including) that event. For example, a vector
clock attached to an event represents the event's causal ``position'' in the
timeline, allowing for causal comparisons between different events. In a sense,
the vector clock---maintained by iterating over the timeline's structure
(\mintinline{rust}|update| for same-node events) and (\mintinline{rust}|merge|
for inter-node causal dependencies)---compresses the whole timeline up to that
event into a single value that captures ``what is essential'' for the purpose
of determining causal relations between events. But what if our goal is
different, \eg, to \emph{summarise} what happened in a timeline in order to
inform our testing? \autoref{fig:example-summary} shows a simplified version of
the default abstraction used by \tool. Instead of accumulating a causal
timestamp like vector clocks do, an \mintinline{rust}|EventHistory| accumulates
for every node in the timeline a set of \emph{happens-before pairs} of event
types that occurred locally at that node. To get an intuition for this,
\autoref{fig:ooda-loop} shows in its centre portion a timeline with its
associated \mintinline{rust}|EventHistory|.\footnote{More precisely, the shown
\mintinline{rust}|EventHistory| is associated with an artificial event that is
causally after every node's last event.}
The abstraction is obtained by
traversing the timeline's events in causal order (starting from an empty
\mintinline{rust}|EventHistory|), and calling \mintinline{rust}|update| for
events on the same node and \mintinline{rust}|merge| for inter-node causal
dependencies. The result is a summary of the timeline's events that captures
some of the history's essential aspects
and which we use in the next stage to
decide which faults to introduce.

\begin{figure}[t]
\setlength{\belowcaptionskip}{-10pt}
\setlength{\abovecaptionskip}{5pt}

\begin{minted}[fontsize=\small, linenos,numbersep=12pt,xleftmargin=20pt]{rust}
pub struct EventHistory {
  events: Map<NodeId, Set<EvKind>>,
  pairs: Map<NodeId, Set<(EvKind, EvKind)>>, 
}
impl TimelineAbstraction for EventHistory {
  fn update(&mut self, ev: &Event) {
    self.events[ev.node].insert(ev.kind);
    for ev_a in self.events {
      let pair = (ev_a, ev.kind);
      self.pairs[ev.node].insert(pair); 
    } 
  }
  fn merge(&mut self, other: &Self, ..) {
    /* For every NodeID, take set union. */ 
  } 
}
\end{minted}

\caption{A timeline abstraction that tracks \emph{happens-before pairs} of
event types on a per-node basis, \eg, a \texttt{Commit} happened before a
\texttt{Rollback} at the same node.}
  
\label{fig:example-summary}
\end{figure}


\subsection{Making Decisions with Q-Learning}
\label{sec:learn}


Equipped with a way to understand the behaviour of the SUT, in the form of timeline abstractions, \tool must decide which actions to take in response to what it observes.

For fuzzing sequential programs, \emph{mutation-based power scheduling} has become the standard approach to generate novel inputs for the program under test based on observations: test inputs that exercise new behaviours are stored and mutated many times to obtain new inputs. However, this technique is ill-suited for testing distributed or reactive systems. The issue is that in the mutation-based paradigm, behavioural feedback is given for the whole input to the system under test (SUT) after execution ends. This is reasonable for
sequential programs, but not for reactive programs---the (temporal and causal) connection between fault introduced and behaviour induced is lost. Indeed, we want our fuzzer itself to be reactive and give behavioural feedback after every action taken rather than only at the end of a long schedule. This complements \jep's \emph{generative fuzzing} approach by giving behavioural feedback after every generated fault and makes \tool adapt in real-time to the SUT. 

The concept of \emph{timeline-driven testing} is key to how \tool adapts to the SUT. To decide which faults to introduce into the SUT, \tool employs Q-learning~\cite{watkins1989learning, watkins1992q}, a model-free reinforcement learning approach. Q-learning enables an agent to dynamically learn a \emph{state-action policy}, \ie, a pairing between the observed state of the environment and the optimal action to take. Based on the observed states, this policy guides the agent in selecting actions to take. After performing an action, the agent receives an immediate reward, which further refines the policy. By associating observed states with optimal actions, the agent maximises the expected rewards over its lifetime.

Q-learning can  be seamlessly integrated into our distributed system fuzzer. With this approach, our fuzzer (\ie, the agent) learns a policy throughout the fuzzing campaign, which serves as a guide to explore diverse states. Specifically, during the fuzz campaign, our fuzzer observes the state $s_i \in \mathcal{S}$ of the SUT, and then employs the learned policy to select a fault $a_i \in \mathcal{A}$ (\ie, an action) to introduce, causing it to transition to the next state $s_{i+1}$. Simultaneously, the fuzzer receives an immediate reward $r_i \leftarrow \mathcal{R}(s_i, a_i, s_{i+1})$. Using the received reward, our fuzzer further refines the policy $\pi: \mathcal{S} \mapsto \mathcal{A}$. Subsequently, our fuzzer selects the next fault $\pi(s)$ to introduce. Over time, our fuzzer progressively learns an optimal policy that maximises the number of distinct states observed (\ie, rewards). In the following, we elaborate this process further.

\paragraph{Capturing States}
In the context of Q-learning, states should represent the behaviour of the environment. As elaborated in \autoref{sec:timelines}, to describe the behaviour of the SUT, we adopt the timeline abstractions (\eg, a set of happens-before pairs). To incorporate Q-learning into our framework, we take a hash of the timeline abstractions to serve as \emph{abstract states}. However, while computing states, we encounter a challenge due to the sensitivity and precision of our timeline abstractions. In order to improve the learning speed of the state-action policy, we aim to treat states that differ only insignificantly as identical. Using a standard hash function for this purpose proves ineffective, as it tends to be overly sensitive to minor variations in the timeline abstraction. Such variations can arise even when the system is operating without any injected faults and under a constant load, owing to the inherent non-deterministic nature of distributed systems.

To address this issue, we adopt MinHash, a locality-sensitive hash function that maps similar input values to similar hash values. In our specific case, to decide whether an observed timeline abstraction corresponds to a new distinct state, we hash it into a signature and then compare this signature to those of previously encountered states. We classify a state as distinct if the similarity falls below a threshold $\varepsilon$. To choose $\varepsilon$, we conduct a calibration stage before fuzzing by running the SUT without any faults and under constant load and observing the timeline abstractions thus obtained. The convention in previous fuzzing works, which we also adopt, is to choose an $\varepsilon$ value that makes 90\% of such ``steady'' states coincide~\cite{stateafl, ankou}.

\paragraph{Learning the Policy}
In our approach, the policy $\pi$ is represented as a Q-table, where each column corresponds to a specific type of fault, and each row represents a distinct state. The available fault types are provided by the fuzzer and enabled at the start of the fuzzing campaign. As new distinct states are observed and added, the rows dynamically increase to accommodate them. Within the Q-table, each cell stores the Q-value for a state-action pair. When a new distinct state is added to the table, the Q-values for each state-action pair associated with that state are initialised to 0. With this setup, the process of refining the policy becomes simplified, involving adjustments to the Q-values.

Q-values are updated in response to rewards. After executing a fault, the fuzzer receives an immediate reward determined by the reward function. The reward function is devised based on our goal. Since the goal is to maximise the number of distinct states observed, we set our reward function to give a constant negative reward (-1) to states that have been observed previously. This approach incentivises \tool to steer the SUT towards unobserved behaviors, promoting exploration. We design the reward function as follows:
\begin{equation}\label{eq:reward_function} 
	\mathcal{R}(s_i, a_i, s_{i+1})=\left\{
	\begin{aligned}
		-1 & , & \textup{if} ~s_{i+1} \in \mathcal{S}_{\mathit{observed}} \\
		0 & ,  & \textup{if} ~s_{i+1} \notin \mathcal{S}_{\mathit{observed}}
	\end{aligned}
	\right.
\end{equation}

Once receiving the rewards, our fuzzer dynamically adjusts Q-values using the Q-learning function $\mathcal{Q}$: $\mathcal{S} \times \mathcal{A} \rightarrow \mathbb{R}$, which determines the Q-value for a specific state-action pair. When an action $a_i$ is executed, it leads to a new state $s_{i+1}$ from the previous state $s_i$ (\ie, $s_i \stackrel{a}{\longrightarrow} s_{i+1}$). Subsequently, we update the Q-value as follows:
\begin{equation}\label{eq:q_function} 
    \mathcal{Q}(s_i, a_i) \leftarrow (1 - \alpha)\mathcal{Q}(s_i, a_i) + \alpha\left(\mathcal{R} + \gamma\mathop{max}\limits_{a^{\prime}}\mathcal{Q}(s_{i+1}, a^{\prime})\right)
\end{equation}

where $\alpha \in (0, 1]$ indicates the learning rate and $\gamma \in (0, 1]$ is a discount factor. The default values chosen for these parameters are $\alpha = 0.1$ and $\gamma = 0.6$, which we determined to work well empirically. With the Q-function, the new Q-value is computed and subsequently updated into the Q-table. As the fuzz campaign proceeds, our fuzzer gradually fine-tunes the Q-values, thus refining the policy to make better decisions.

\paragraph{Getting Next Fault}
When the SUT is in a state $s$, our fuzzer selects the next action $a$ based on the learned policy. In the current state $s$, we first obtain the Q-values for all $n$ actions and then utilise the softmax function to convert these Q-values into a probability distribution $\mathcal{D}$. The probability $\mathcal{D}(i)$ for each action $a_i$ ($i \in n$) is calculated as follows:
\begin{equation}\label{eq:softmax} 
    \mathcal{D}(i) \leftarrow \frac{e^{\mathcal{Q}(s, a_i)}}{\sum\nolimits_{j=1}^{n}e^{\mathcal{Q}(s, a_j)}}
\end{equation}

To decide the next action to take, we sample from the probability distribution $\mathcal{D}$. To achieve this, we generate a random number $p \in [0, 1]$, and then check the cumulative probability of $\mathcal{D}(i)$ until we identify the first action where the cumulative probability exceeds $p$. This action is chosen as the next action to execute. Actions with higher probabilities have a greater likelihood of being chosen, while actions with lower probabilities still have a chance of being selected. This approach ensures a balance between favoring actions with higher probabilities while maintaining the possibility of choosing actions with lower probabilities.

Thus far, we have introduced each individual component of \autoref{algo:jeppa}. To kickstart the fuzzing campaign, we set the default maximum steps as 12, each with a $2.5$-second window. Following each schedule, there is a 5-second period allocated to reset the system to a stable state.

%% file: evaluation.tex
\section{Evaluation}
\label{sec:evaluation}

We implement \tool on top of \jepsen-0.2.7, to test distributed system
implementations written in C, C++, and Rust. To enable code-level
instrumentation, we created a LLVM compiler pass (similar to that used by
AFL~\cite{afl}) to add into the compiled binary our event instrumentation, as
described in \autoref{sec:event3}. The code implementing this pass measures
roughly 1,000 lines of C/C++ code. The observers at each node that collect
events and the mediator which collates events from all observers, intercepts
packets, constructs and abstracts timelines, and learns the policy required to
guide fault injection, are implemented in Rust. The code for these components
measures roughly 9,000 lines of Rust code. To enact faults, we implemented a
linker in \jep that asks \tool for the next fault to execute. This linker
consists of 140 lines of Clojure code.

\subsection{Evaluation Setup}

To evaluate the effectiveness and efficiency of \tool in exploring distinct
program behaviours and finding bugs in industrial distributed system
implementations, we have designed experiments to address the following
questions:

\begin{description}
\item [\textbf{RQ1}] \textbf{Coverage achieved by \toolBold.} Can \tool cover more distinct system states than \jepsen? 
\item [\textbf{RQ2}] \textbf{Efficiency of bug finding.} Can \tool find bugs more efficiently than \jepsen?
\item [\textbf{RQ3}] \textbf{Discovering new bugs.} Can \tool discover new bugs in rigorously-tested distributed system implementations?
\end{description}

\subsubsection{Baseline tool.} 

We selected \jepsen as our baseline tool due to its popularity in
stress-testing distributed system implementations. To our knowledge, \jepsen is
the only widely-used black-box fuzzer in this domain. It has gained recognition
for its user-friendliness and has helped to uncover numerous bugs in real-world
implementations of distributed systems. By building on top of \jepsen, we have
developed \tool to enhance the effectiveness of fuzzing while preserving
\jepsen's ease of use. Another black-box fuzzer, called \namazu~\cite{namazu},
is less popular and can only test Go/Java programs.

As described in the introduction, white-box fuzzers such as
\modist~\cite{YangCWXLLYLZZ09} and \flymc~\cite{LukmanKSSKSPTYL19} require an
extensive manually-written test harness or heavy deterministic control at the
system level, and are used for systematic testing as opposed to stress-testing.
Due to their heavy-weight nature, they target a different use case compared to
\tool and are less practical to adopt in industry.

\subsubsection{Subject programs.} \autoref{tab:subjects} presents the subject
programs included in our evaluation. It consists of six open-source distributed
system implementations written in C, C++, and Rust. We selected these subjects
because: (1) they are widely used in the industry, (2) they can be instrumented
by our LLVM pass, and (3) they have undergone rigorous testing using \jepsen
either by contracting \jepsen's author\footnote{Test reports are public at
{\url{https://jepsen.io/analyses}}} or by rolling their own \jepsen test
harness. Finding new bugs in these systems would be a strong indication that
\tool performs better than \jepsen.

\input{tables/subjects}

\subsubsection{Event Annotations.} 
To instrument code events, we annotated a total of 103 to 157 functions and basic blocks in our subject programs using \ccode{//INSTRUMENT_FUNC} and \ccode{//INSTRUMENT_BLOCK} forms. Specifically, we made 120 annotations for Braft, 108 annotations for Dqlite, 103 annotations for MongoDB, 129 annotations for Redis, 157 annotations for ScyllaDB, and 138 annotations for TiKV. This process required one of our authors to dedicate a total of 2 hours, averaging approximately 20 minutes for each subject. 

\begin{figure}[t]
\centering
\begin{minted}[fontsize=\footnotesize,xleftmargin=0pt,]{c}
145    int membershipRollback(struct raft *r){
146      ...
158      // Fetch the last committed configuration entry 
159      entry = logGet(&r->log, r->config_index);
160      assert(entry != NULL);
176    }

985    // INSTRUMENT_FUNC
986    static int deleteConflictingEntries(){
987      ...
1007     // Possibly discard uncommitted config changes
1008     if (uncommitted_config_index >= entry_index){
1009       // INSTRUMENT_BLOCK
1010       rv = membershipRollback(r); 
1011     }
1043   }
\end{minted}
\caption{Annotating \dqlite for membership rollback.}
\label{fig:annotation}
\end{figure}

We annotate ``interesting" code events primarily based on the Raft and Paxos TLA+ specifications.\footnote{The specifications are taken from {\url{https://github.com/tlaplus/Examples}}.} We now demonstrate the annotation process for events in the motivation example (\autoref{fig:rollback}), and present the annotated code in \autoref{fig:annotation}. In the Raft TLA+ specification, one event involves the removal of conflict entries. To annotate this event, we search for the ``conflict" keyword in the \dqlite source code, resulting in 7 matched locations. However, among these matches, only one location pertains to functions or basic blocks, specifically the ``deleteConflictingEntries" function. Therefore, we add the annotation \ccode{//INSTRUMENT_FUNC} before this function. Instead of annotating a function - we can also decide to annotate (some of) the basic blocks representing the calling locations of the function. However, this may require additional work in terms of determining which calling location to instrument. For example, for the "membershipRollback" function, we can annotate the basic block in line 1009 that invokes this function with the annotation \ccode{//INSTRUMENT_BLOCK} shown in \autoref{fig:annotation}. But choosing this calling location, takes into consideration the condition (line 1008) that there must be an uncommitted configuration entry among the conflicting entries to be removed (explained in \autoref{sec:raft}). Once the code is annotated, our compile-time instrumentation automatically instruments the annotated code without any manual effort.
\paragraph{Discussion}
It is worth mentioning that we adopt the heuristic based on TLA+ specifications to instrument events in our evaluation. Nevertheless, users of \tool have the flexibility to instrument their own custom event types in accordance with their own heuristics. For instance, this might involve the instrumentation of error handlers \cite{YuanLZRZZJS14} or \emph{enum} cases \cite{sgfuzz}. The identification of such events can be achieved fully automatically, and we have integrated this functionality into the \tool tool. 


\subsubsection{Configuration parameters.} 

To distinguish between distinct
states, we set the similarity threshold $\varepsilon$ to 0.70. To detect bugs,
we adopt several test oracles: (1) AddressSanitizer (\ie, ASan) for exposing
memory issues, (2) log checker to detect issues in application logs by scanning
for keywords such as ``\emph{fatal}'', ``\emph{error}'' and ``\emph{bug}'', and
(3) consistency checker \elle~\cite{elle} to check consistency violations. We
set up the same number of nodes as the existing \jepsen tests: 9 nodes
for MongoDB and 5 nodes for the other subjects under test. To ensure a fair
comparison, we enabled the same faults in \tool as those used in the original
\jepsen tests (\ie, our tool does not have access to more fault types).

All experiments were conducted on Amazon Web Services using the m6a.4xlarge
instance type. This instance type has 64 GB RAM and 16 vCPUs running a 64-bit
Ubuntu TLS 20.04 operating system. Following community suggestions, we ran each
tool for 24 hours and repeated each experiment 10 times. To reduce statistical
errors, we report as results the average values obtained over the 10 runs.

\subsection{Coverage Achieved by \toolBold (RQ1)}

For the first experiment, we monitor the number of distinct states (see
\autoref{sec:learn} for the definition of states) exercised by \tool and
\jepsen over time, and compare their state coverage capabilities.

\begin{figure}[t]
    \centering
    \vspace{.3cm}
    \includegraphics[page=1, trim= 0.4in 1in 0.4in 1.8in, clip, scale=0.38]{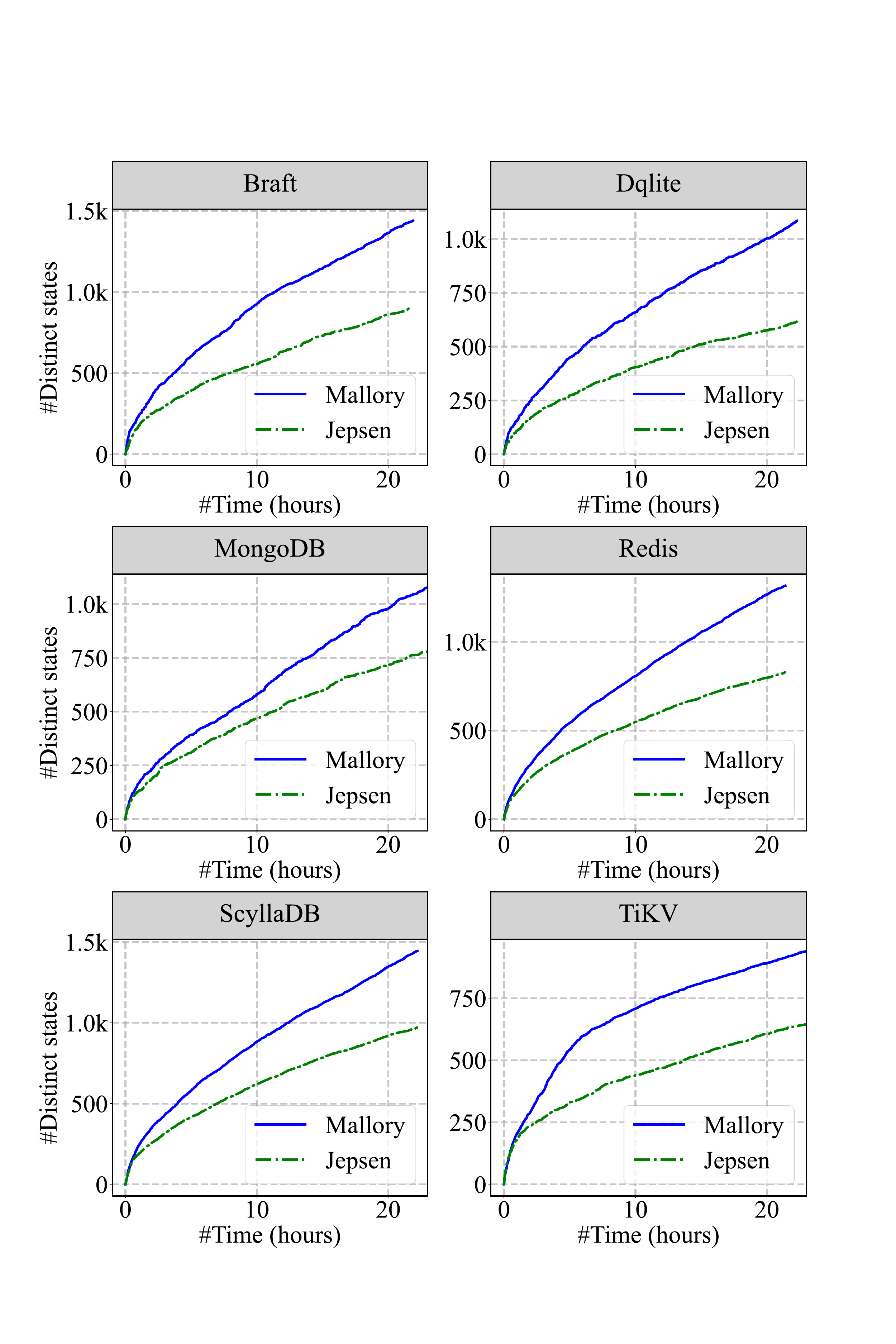}
    \caption{The trends in the average number of distinct states within 24 hours across 10 runs.}
    \label{fig:state-trend}
    \vspace{-.3cm}
\end{figure}

To observe states exercised by \jepsen, we ran \jepsen in the same setup as \tool, but without controlling the fault injection. We collected the average number of distinct states achieved by \tool and \jepsen within 24 hours across 10 runs, and we present the comparison in~\autoref{fig:state-trend}. As shown in this figure, \tool outperforms \jepsen by covering more distinct states in the same time budget, thus exercising the SUT under more diverse scenarios. 

Initially, at the start of each experiment, the number of distinct states achieved by \tool is similar to that achieved by \jepsen, 
which is expected. \tool utilises Q-learning to learn an optimal state-action policy during fuzzing, guiding it to make better decisions about actions to take in observed states. However, during the initial phase of fuzzing, this policy is not yet well-learned and lacks sufficient knowledge to avoid exploring redundant states. Consequently, both \tool and \jepsen struggle to select the optimal actions for the observed states. 

However, as fuzzing progresses, \tool gradually refines the policy by assigning negative rewards to certain state-action pairs, thereby disincentivizing certain actions selected in particular states. This disincentive effectively curtails the exploration of repetitive states and steers \tool towards exploring more diverse states. The policy learned using Q-learning proves to be highly effective. As evident in \autoref{fig:state-trend}, over time, the number of distinct states covered by \tool is significantly more than that covered by \jepsen. We do not observe the number of distinct states saturating in either tool, but \tool's exploration rate of distinct states is higher than that of \jepsen. Additionally, the gap between \tool and \jepsen consistently widens, indicating that the Q-learning policy continuously improves and becomes increasingly effective in guiding \tool's exploration process.


The state coverage statistics of \tool over \jepsen are listed
in~\autoref{tab:statistic}. The ``State-impr'' column shows the average
improvement of \tool in the number of distinct states at the end of 24 hours,
over 10 runs. Our results show that \tool covers an average of 54.27\% more
distinct states than \jepsen on our test subjects, with an improvement ranging
from 36.48\% to 76.14\%. The ``Speed-up'' column indicates the average speed-up
of \tool compared to \jepsen in achieving the same number of observed states.
On average, \tool archives a 2.24$\times$ speed-up over \jepsen. To mitigate
the effect of randomness, we measured the Vargha-Delaney (${\hat{A}_{12}}$) and
Wilcoxon rank-sum test ($\textit{U}$) of \tool against \jepsen.
${\hat{A}_{12}}$ is a non-parametric measure of effect size that provides the
probability that random testing of \tool is better than random testing of
\jepsen. $\textit{U}$ is a non-parametric statistical hypothesis test that
determines whether the number of distinct states differs across \tool and
\jepsen. We reject the null hypothesis if $\textit{U} \textless 0.05$,
indicating that \tool outperforms \jepsen with statistical significance. For
all subjects, ${\hat{A}_{12}} = 1$ and $\textit{U} \textless 0.01$ for \tool
against \jepsen. This demonstrates that \tool significantly outperforms
\jepsen.

\input{tables/statistic}

Furthermore, we measured the memory consumption required to maintain the data
structure of the Lamport-style timeline. The average memory consumption was 3.21
GB, which we consider acceptable. Memory consumption remains stable over time,
as we only retain the portion of the timeline needed for abstraction and remove
already-abstracted portions. Additionally, our fuzzer is designed to learn and
react to observations in real time (see \autoref{sec:learn}). We measured the
time taken from the point of fault injection to receiving the behaviour
feedback and found that in 92.20\% of cases, this process took less than one
window time, \ie,~\tool receives feedback and adapts its policy before it has to
decide the next action.

\result{In terms of state exploration, \tool covers 54.27\% more distinct
states than \jepsen with a 2.24$\times$ speed-up.}

\subsection{Efficiency of Bug Finding (RQ2)}
\label{sec:reproducing}

To evaluate the efficiency of \tool at finding bugs, we compared \tool
and \jepsen with regards to the time required to reproduce existing
bugs. To this end, we created a dataset of bugs by selecting 10 recent
issues from each subject's GitHub issue list (from early 2019 to April
2023) that contained instructions for reproduction. We attempted to
reproduce the bugs manually and included any successfully reproduced
bugs in our dataset. We finally collected a total of 16 bugs across
all subjects. The bug IDs and types of bugs are presented in the first
two columns of \autoref{tab:reproduction_results_short}.

We ran both tools, \tool and \jepsen, on buggy versions of the subjects for 24
hours, repeated 10 times. The last main column shows the time used for each
tool to expose the bug. We marked ``T/O'' if one tool failed to find the bug
within the given time budget. Overall, in these 16 bugs, \tool successfully
exposed 14 bugs, while \jepsen only found 9 bugs. In terms of time usage, \tool
takes much less time (\ie, 6.13 hours on average), while \jepsen needs 11.45
hours. Hence, compared with \jepsen, \tool achieves a speed-up of 1.87$\times$
in bug finding.

For shallow bugs whose states are easy to reach, such as Dqlite-338 and
Dqlite-327, \tool and \jepsen perform well and perform similarly. However, for
deep bugs that are harder to expose, \tool performs much better than \jepsen.
For example, to expose Redis-51, \jepsen took 6.40 hours, while \tool only took
1.66 hours. This is attributed to a faster state-exploration speed of \tool. In
addition, since \tool explored more distinct states than \jepsen, \tool was
also able to expose more bugs. Specifically, \tool successfully exposed
Dqlite-324, Dqlite-323, Redis-28, Redis-23, and Redis-17, while \jepsen had
difficulty in exposing them. We further investigated the two bugs (\ie,
Dqlite-356 and Dqlite-314) missed by \tool, and found exposing these bugs needs
to inject specific environment faults (\eg, disk faults), which were not
included in our evaluation. 

To mitigate randomness, we adopt the Vargha-Delaney (${\hat{A}_{12}}$) to
measure the statistical significance of performance gain. The last subcolumn of
\autoref{tab:reproduction_results_short} shows these results. We mark
${\hat{A}_{12}}$ values in bold if they are statistically significant (taking
0.6 as a significant level, or threshold). We can see that, in most cases,
\tool significantly outperformed \jepsen.

\result{In terms of bug finding, \tool finds 5 more bugs and finds bugs
1.87$\times$ faster than \jepsen.}

\input{tables/reproducing-short}


\input{tables/new_bugs}

\subsection{Capability of Exposing New Bugs (RQ3)}

To evaluate \tool's capability of exposing new bugs, we utilised \tool on the
latest versions of our subjects. In the course of the experiment, \tool
produced promising results, as demonstrated in \autoref{tab:new_bugs}. Although
all of these subjects have been rigorously tested by \jepsen and other tests,
\tool was still able to find a total of 22 previously unknown bugs, and 18 bugs
of them were confirmed by their developers. Out of these 22 bugs, 10 bugs were
associated with vulnerabilities, and we have requested CVE IDs for them. As of
the paper submission, we have already obtained 6 CVE IDs and the remaining
requests are still being processed.


We conducted a thorough analysis of the nature of these new bugs found
by \tool, shown in~\autoref{tab:new_bugs}. The table also includes the
bug checkers used to uncover these bugs. Among these 22 bugs, 7 bugs
were determined to be consistency violations exposed by the \elle
consistency checker. Three bugs (\#3, \#4, and \#17) violated the Raft
protocol due to missing leaders or the existence of two leaders in the
same term, and they were detected by the log checker. AddressSantizer
(\ie, ASan) exposed 5 memory issues. Furthermore, the log checker
detected 7 runtime failures or invariant violations. These results
indicate that \tool is beneficial to expose diverse types of
previously unknown bugs.

In addition, we applied \jepsen to detect these 22 new bugs; however,
under the allotted time limit, \jepsen was only able to detect four of
them (\ie, the bugs \#1, \#7, \#16, and \#22), as shown in the last
column of \autoref{tab:new_bugs}. This result is expected because
these subject systems routinely undergo \jepsen testing by their
developers, making it challenging for \jepsen to discover new bugs.

In the following, we provide two case studies to illustrate bugs that were
exposed by \tool.

\vspace{3pt}

\paragraph{\emph{Case study: Bug \#2 in Braft}} \braft is a robust Raft
implementation designed for industrial applications, which is widely used
within Baidu to construct highly available distributed systems. However, a
critical vulnerability, known as Bug \#2, remained undetected in all \braft
release versions from 2019 until its recent patch. This bug occurs when a
server cannot release its allocated memory before failure, resulting in a
memory leak issue.

To trigger this issue, a minimum of three environmental faults must be
introduced sequentially. Initially, the server dynamically allocates 
enough memory for its operation, which is explicitly managed by itself. However, before releasing the 
allocated memory, the server is paused, and its memory remains in use. Subsequently, the server is resumed,
only to become coincidentally isolated from the cluster due to a network
partition, resulting in a failed start. Hence, the server crashes without the
chance to release the memory allocated. This bug happens due to a flawed
logic design; specifically, the allocated memory can only be released when the
process is in the running status, and the allocated memory cannot be released
before running. This logic design is reasonable in stable environments without
faults, as only the running server may have allocated memory. However, in this
extreme environment, the shortcoming in the logic is exposed.

Bug \#8 of \dqlite is another memory leak issue, similar to Bug~\#2 in \braft. It
remained hidden in \dqlite for approximately four years and affected all its
release versions before we found it. Bug \#2 in \braft and Bug \#8 in \dqlite
both evaded the rigorous testing efforts by their developers, demonstrating how
our tool \tool can significantly reduce the exposure of systems to
vulnerabilities.

\vspace{3pt}

\paragraph{\emph{Case study: Bug \#11 in \dqlite}} Although \jepsen is already
part of \dqlite's Continuous Integration process, \tool has managed to
expose several new bugs in \dqlite. The developers have expressed a keen
interest in \tool and are awaiting its open-source release so that they can
incorporate it into their testing. 

Bug~\#11 in \dqlite is caused by the snapshotting of uncommitted logs,
and it is reminiscent of the membership rollback bug shown in
\autoref{fig:timeline}. The schedules required to trigger these bugs
are quite similar, but Bug \#11 is not triggered by the configuration
change. Specifically, the event \ding{172} involves one plain
read/write log entry instead of the configuration change. After this
event, the cluster undergoes the same sequence of environment faults,
including a network partition $(\#\!\set{S_1, S_2}$,
$\#\!\set{S_3, S_4, S_5})$, leader $S_1$ crashing, and network
healing. As a result of these faults, server $S_2$ ends up with
conflicting entries with the leader $S_3$, which must be removed.
However, the conflicting entries are in a snapshot, which makes the
removal fail. This failure leads to the server becoming unavailable.
Although the schedule to trigger Bug \#11 is slightly shorter than that of the membership rollback bug in \autoref{fig:timeline}, 
\jepsen, which adopts a random search strategy, failed to expose it during our experiments within the allotted time. In contrast, \tool, guided by the policy learned with Q-learning, successfully exposed this bug. This is easily explained considering the number of states explored by \jepsen and \tool shown in \autoref{fig:state-trend}. Within 24 hours, \jepsen only covered 600 states, while \tool explored over 1,000 states. With the capability to explore more novel behaviors, \tool significantly increased its chances of exposing bugs. This instance demonstrates the effectiveness of the policy learned using Q-learning in guiding \tool to explore more behaviors and, consequently, enhance the likelihood of bug exposure.


\result{\tool found 22 zero-day bugs in rigorously tested implementations, and
18 bugs out of them have been confirmed by their developers. 10 of these 18
bugs correspond to security vulnerabilities, and out of these 6 CVEs have been
assigned.}

%% file: tables/subjects.tex
\begin{table}[t]
    \small
    \setlength{\abovecaptionskip}{5pt}%
    \setlength{\belowcaptionskip}{0pt}%
    \setlength\tabcolsep{0.7pt}
    \caption{Detailed information about our subject programs.}
    \label{tab:subjects}
    \def\arraystretch{1.1}
    \begin{tabular}{l|ll|lrr}
        \toprule
        {\bfseries Subject} & {\bfseries Description} & {\bfseries Protocol} & {\bfseries Lang.} & {\bfseries \#LOC} & {\bfseries \#Stars} \\
        \hline
        \hline
        Braft     & Baidu Raft implementation & Raft &	C++ &  31.6k  &  3.5k\\
        Dqlite    & Distributed SQL DBMS & Raft &	C   & 54.2k   &  3.4k \\
        MongoDB   & Distributed NoSQL DBMS & Raft &	C++ &  1121.6k  &  23.6k\\
        Redis     &	Distributed in-memory DBMS & Raft &   C   & 211.4k   &  59.6k \\
        ScyllaDB  & Distributed NoSQL DBMS & Raft/Paxos &	C++ &  122.4k  &  9.8k \\
        TiKV      & Distributed key-value DBMS  & Raft &	Rust&  404.5k   &  13.0k\\
        \bottomrule
    \end{tabular}
\end{table}

%% file: tables/statistic.tex
\begin{table}[t]
    \setlength{\abovecaptionskip}{5pt}%
    \setlength{\belowcaptionskip}{0pt}%
    \caption{Statistics of distinct state numbers achieved by \toolBold compared to that achieved by \jep.}
    \label{tab:statistic}
    \small
    \setlength\tabcolsep{4pt}
    \def\arraystretch{1.1}
    \begin{tabular}{l|cc|cc}
        \toprule
        {\bfseries Subject} & {\bfseries State-impr} & {\bfseries Speed-up} & ${\boldsymbol{\hat{A}_{12}}}$ & {\bfseries \textit{U}} \\
        \hline
        \hline
        Braft   & 59.34\% &	2.28$\times$ &	1.00 &  \textless 0.01\\
        Dqlite & 76.14\%  &	2.56$\times$ &	1.00 &  \textless 0.01 \\
        MongoDB & 36.48\% &	1.57$\times$ &	1.00 &  \textless 0.01 \\
        Redis & 58.92\%	 & 2.06$\times$ &	1.00 &  \textless 0.01\\
        ScyllaDB & 48.82\% &	1.88$\times$	& 1.00 &  \textless 0.01\\
        TiKV & 45.93\%	& 3.07$\times$	& 1.00 &  \textless 0.01 \\
        \hline
        \hline
        {\bfseries AVG} & 54.27\% &	2.24$\times$ & - & - \\
        \bottomrule
    \end{tabular}
\end{table}

%% file: tables/reproducing-short.tex
\begin{table}[t]
    \setlength{\abovecaptionskip}{5pt}%
    \setlength{\belowcaptionskip}{0pt}%
    \caption{Statistics of reproduced known bugs and the performance
      of both \toolBold and \jepsen in exposing these bugs.}
    \label{tab:reproduction_results_short}
    \small
    \setlength\tabcolsep{4pt}
    \def\arraystretch{1.1}
    \begin{threeparttable}
    \begin{tabular}{l|l|ccc}
        \toprule
        \multirow{2}{*}{\bfseries Bug ID} & \multirow{2}{*}{\bfseries Type of bug} & \multicolumn{3}{c}{ \bfseries Time to exposure}  \\ 
        \cline{3-5} 
         &  & {\footnotesize \toolBold} & {\footnotesize \jepsenBold} &  \footnotesize {${\boldsymbol{\hat{A}_{12}}}$} \\
        \hline
        \hline
        Dqlite-416 & Null pointer deference & \textbf{0.76h} & 1.44h & \textbf{1.00}\\
        Dqlite-356 & Snapshot installing failure & T/O & T/O & 0.50\\
        Dqlite-338 & Election fatal with split votes & 0.16h & 0.16h & 0.50 \\
        Dqlite-327 & Member removal failure & 0.06h & 0.05h & 0.49\\
        Dqlite-324 & Log truncation failure & \textbf{5.94h} & T/O & \textbf{1.00}\\
        Dqlite-323 & Membership rollback failure & \textbf{8.68h} & T/O & 
        \textbf{1.00}\\
        Dqlite-314 & Crashing on disk failure & T/O & T/O & 0.50\\
        \hline
        Redis-54 & Snapshot panic & \textbf{3.33h} & 5.00h & \textbf{0.95}\\
        Redis-53 & Committed entry conflicting & \textbf{0.87h} & 1.17h & \textbf{0.89}\\
        Redis-51 & Not handling unknown node & \textbf{1.66h} & 6.40h & \textbf{1.00}\\
        Redis-44 & Loss of committed write logs & \textbf{0.34h} & 0.58h & \textbf{0.60}\\
        Redis-43 & Snapshot index mismatch & 0.16h & 0.16h & 0.50\\
        Redis-42 & Snapshot rollback failure & 0.29h & 0.26h & 0.50\\
        Redis-28 & Split brain after node removal & \textbf{9.56h} & T/O & \textbf{1.00}\\
        Redis-23 & Aborted read with no leader & \textbf{7.29h} & T/O & \textbf{1.00}\\
        Redis-17 & Split brain and update loss & \textbf{11.06h} & T/O & \textbf{1.00}\\
        \hline
        \hline
        \multicolumn{2}{l|}{ Bugs exposed in total} & 14 & 9 & -\\
        \hline
        \multicolumn{2}{l|}{ Average time usage} & 6.13h & 11.45h & - \\
        \hline
        \multicolumn{2}{l|}{ Speed-up on time usage} & - &  1.87$\times$ & -\\
        \bottomrule
        \end{tabular}
        \begin{tablenotes}
            \footnotesize
            \item[1] T/O means that the tool cannot expose bugs within 24 hours for 10 experimental runs. We replace T/O with 24 hours when calculating average usage time.
            \item[2] Statistically significant values of $\hat{A}_{12}$ are shown in bold.
        \end{tablenotes}
    \end{threeparttable}
    \vspace{-.3cm}
\end{table}

%% file: tables/new_bugs.tex
\begin{table*}[]
    \setlength{\abovecaptionskip}{5pt}%
    \setlength{\belowcaptionskip}{0pt}%
    \caption{Statistics of the zero-day bugs discovered by \toolBold in rigorously tested systems; a total of 22 previously unknown bugs, 18 bugs confirmed by their developers, and 10 software vulnerabilities. }
    \label{tab:new_bugs}
    \small
    \setlength\tabcolsep{4pt}
    \def\arraystretch{1.1}
    \begin{tabular}{r|l|lll|c}
        \toprule
        {\bfseries ID} & {\bfseries Subject} & {\bfseries Bug description} & {\bfseries Bug checker} & {\bfseries Bug status} & {\bfseries \jepsenBold?} \\
        \hline
        \hline
        1 & Braft & Read stale data after a newly written update is visible to others & \elle & Investigating & \cmark \\
        2 & Braft & Leak memory of the server when killed before its status becomes running  & ASan & CVE-Granted, fixed & \xmark\\
        \hline
        3 & Dqlite & Two leaders are elected at the same term due to split votes & Log checker & Confirmed & \xmark \\
        4 & Dqlite & No leader is elected in a healthy cluster with an even number of nodes & Log checker & Confirmed, fixed & \xmark \\
        5 & Dqlite & A node reads dirty data that is modified but not committed by another node & \elle & Confirmed & \xmark \\
        6 & Dqlite & Lose write updates due to split brain & \elle & Confirmed & \xmark \\
        7 & Dqlite & A null pointer is dereferenced due to missing the pending configuration & ASan & CVE-Requested & \cmark \\
        8 & Dqlite & Leak allocated memory when failing to extend entries & ASan & CVE-Requested, fixed & \xmark \\
        9 & Dqlite & Buffer overflow happens while restoring a snapshot  & ASan & CVE-Requested & \xmark \\
        10 & Dqlite & A node has an extra online spare  &  Log checker & Confirmed & \xmark\\
        11 & Dqlite & Violate invariant as a segment cannot open while truncating inconsistent logs  &  Log checker & CVE-Requested & \xmark\\
        \hline 
        12 & MongoDB & Not repeatable read due to missing the local write update & \elle & Confirmed & \xmark\\
        13 & MongoDB & Not read committed due to missing the newly written update & \elle & Confirmed & \xmark \\
        \hline 
        14 & Redis & Read stale data after new data is written to the same key & \elle & Confirmed & \xmark \\
        15 & Redis & Buffer overflow due to writing data to a wrong data structure & ASan & CVE-Granted, fixed & \xmark \\
        16 & Redis & Runtime panic on initializing a cluster due to database version mismatch & Log checker & CVE-Granted & \cmark \\
        \hline
        17 & TiKV & No leader is elected for a long time in a healthy cluster & Log checker & Investigating & \xmark \\
        18 & TiKV & Lose write updates due to split brain & \elle & Investigating & \xmark\\
        19 & TiKV & Runtime fatal error when one server cannot get context before the deadline & Log checker & CVE-Granted & \xmark \\
        20 & TiKV & Runtime fatal error in a server when the placement driver is killed & Log checker & CVE-Granted & \xmark \\
        21 & TiKV & Runtime fatal error when failing to update max timestamp for the region & Log checker & CVE-Granted & \xmark \\
        22 & TiKV & Monotonic time jumps back at runtime & Log checker & Investigating & \cmark \\
        \bottomrule
        \end{tabular}
\end{table*}

%% file: related.tex
\section{Related Work}
\label{sec:related}

\paragraph{\emph{General Greybox fuzzing}}
The vast majority of existing grey-box fuzzers aim at testing
sequential software systems, with most of the recent research efforts
dedicated to generating more diverse
inputs~\cite{SteinhofelZ22,formatfuzzer}, defining better feedback
functions~\cite{BaBMR22,vuzzer} and test oracles~\cite{MengDLBR22}.
With this mindset, fuzzing distributed systems poses unique challenges
since ($i$)~the inputs include not just plain data but also schedules
consisting of environmental faults and ($ii$)~code coverage is not as
efficient, since distributed systems typically do not have complex
control flow and their behavioural complexity comes from the
asynchrony of operations across multiple nodes.
The recent \muzz~\cite{muzz}  framework for fuzzing of (single-node) multi-threaded
programs, addresses ($i$)-($ii$) by extending the edge coverage metric
with possible thread interleavings, while also identifying equivalent
schedules.
\muzz' approach does not extend to distributed systems as it is
tailored to tracking the ordering of specific threading functions,
while \tool works with arbitrary events and communication patterns.
Furthermore, \muzz relies on instrumenting the system scheduler, which
is difficult to implement for a distributed system without
virtualising the networking layer. 
Moreover, our timeline-based approach offers a more general way to
observe program behaviours for \emph{any} non-sequential program,
including stand-alone multithreaded as well as distributed systems.

\vspace{2pt}
\paragraph{\emph{Greybox fuzzing of protocols}}
Recently, greybox fuzzing has been extended to stateful reactive systems. One of the key challenges in this research is state identification - since there are many program variables in an implementation, it is hard to gauge the state variables. \aflnet \cite{aflnet} is one of the early works in this domain. It uses response code as a proxy for states, and mutates sequences of protocol messages based on coverage and other feedback. However, response codes may not be an accurate proxy for stateful behavior. Among other works, \stateafl \cite{stateafl} utilises the program's in-memory state to represent the service state. \ijon \cite{ijon} provides motivational examples to show that for some examples it may be intuitive to provide manual annotations to guide fuzzing of stateful software systems. \sgfuzz~\cite{BaBMR22} does not need manual annotations, and instead of simple heuristics to identify state variables such as postulating that state variables are often captured by {\em enum} type variables. Our work on distributed system fuzzing does not make any such assumptions about state variables. 
It captures the events executed so far in a reactive system via a timeline abstraction. 
It also suggests for the first time, the fuzzer of stateful reactive systems, itself as a reactive system. Thus the fuzzer feedback instead of being given for an entire schedule or execution, is given incrementally action by action, with the probabilities for the actions being adjusted via Q-learning.

\vspace{2pt}
\paragraph{\emph{Testing}}
%
The majority of state-of-the-art testing frameworks that explore
behaviours of a distributed system assume \emph{full control} over the
inherent non-determinism of runtime executions, with \jep being a
notable exception~\cite{jepsen}.
They achieve this by either (a)~replacing the networking
layer~\cite{YangCWXLLYLZZ09}, (b)~explicitly modifying SUT to include
a test
harness~\cite{ZhouXSNMTABSLRD21,LeesatapornwongsaHJLG14,OzkanMNBW18},
or (c)~implementing the system in a testing-friendly
language~\cite{KillianABJV07,YuanY20,DesaiPQS18}.
Controlling the asynchrony makes it possible to employ techniques from
software model checking such as \emph{partial order
  reduction}~\cite{JhalaM09} to avoid redundancy when exhaustively
exploring the space of bounded runtime
executions~\cite{OzkanMO19,DragoiEOMN20,OzkanMNBW18,LukmanKSSKSPTYL19}.
While these approaches allow for more effective behaviour exploration
than \jep/\tool, they are far more difficult to apply, requiring,
correspondingly, (a)~specific OS setup, (b)~protocol-aware SUT
modifications, or (c)~using a domain-specific language.

\vspace{2pt}
\paragraph{\emph{Model Checking}}
Of the existing model-checking frameworks,
\modist~\cite{YangCWXLLYLZZ09} is the most similar to our approach, as
it requires no modification in SUT code; instead, it manipulates the
system's execution by intercepting calls to the Windows API.
The main conceptual distinction of \modist from our work is its
interposition layer: unlike our observers, which are passive, \modist
intercepts network, timing, and disk-related system calls and
\emph{pauses} the SUT.
This provides more control---in our terminology, it allows the
mediator to act as a \emph{scheduler}---but comes at the expense of
requiring a complex interposition framework that replicates and
replaces most of the API of a specific OS.
The design innovations of \tool that enable summarising observations
about SUT (\autoref{sec:timelines}) are orthogonal to the use of an
interposition layer, and, therefore,
such an interposition layer could be integrated within our
architecture---rather than choosing only which faults to introduce,
our nemesis would choose every action. In this work, we focused on finding bugs in non-Byzantine
fault-tolerant distributed systems~\cite{LamportSP82}
thus, side-stepping the challenge of 
modelling the behaviour of possibly malicious nodes.
We believe that \tool's workflow can be combined with existing
techniques for Byzantine system testing that emulate
attacks by running several copies of the same node, but, for now, only
allow for execution in a network emulator~\cite{BanoSCPLCM21}.
%



\vspace{2pt}
\paragraph{\emph{Deductive verification}}
%
Finally one can employ a mix of 
algorithmic and deductive logical reasoning, albeit not on actual protocol implementations. These approaches broadly fall into one of the following two lines of
work. The first line of work is concerned with sound verification of
\emph{abstract models} of distributed
protocols~\cite{LamportM94,0001G0LTW22} for safety and liveness
properties and focuses on methods for automated reasoning, such as
checking and inferring protocol
\emph{invariants}~\cite{Padon-al:PLDI16}.
Even though some of those approaches allow one to generate
executable code from verified protocol
models~\cite{TaubeLMPSSWW18,Sergey-al:POPL18}, the implementations tend
to evolve over time, losing their correspondence to the formally
verified models and, hence, relying on testing for correctness.
The second line of work concerns verification of \emph{executable}
code and typically requires the system to be implemented in a
domain-specific language that allows for machine-assisted formal
reasoning~\cite{HawblitzelHKLPR15}.
Such verified implementations incur very high maintenance
costs~\cite{WoosWATEA16} and still remain prone to bugs due to
occasional inadequate \emph{assumptions} about the 
networking infrastructure or third-party
libraries~\cite{FonsecaZWK17}.


%% file: conclusion.tex

\section{Concluding Remarks}
\label{sec:conclusion}


In this work we proposed \tool---the
first adaptive greybox fuzzer for distributed systems.
The key insight behind \tool's design is to summarise the runtime
behaviour of the distributed system under test in the form of
Lamport-style \emph{timelines} that capture causality of events, and
use the timelines to define a feedback function for guiding the search
for bugs.

Our conceptual contribution of \emph{timeline-driven testing} opens
new avenues for automated testing of distributed systems, similar to
what was achieved for sequential programs by tools like
\aflfast~\cite{AFLFast}. 
\aflfast achieves high behavioural diversity by making smart online
decisions about covered program paths, during the fuzz campaign.
Similarly, \tool achieves behavioural diversity by making online
decisions to detect and prioritise novel event sequences that have not
been observed before.

We evaluated \tool on six widely-used and rigorously-tested industrial
distributed system implementations such as \dqlite and \redis. The
experimental results show the effectiveness and efficiency of Mallory
in achieving significantly higher state coverage and faster
bug-finding speed than the state-of-the-art tool \jep. Finally, \tool
discovered 22~previously unknown bugs (10~new vulnerabilities amongs
them) which have contributed to new CVEs.